\documentclass[review,sort&compress]{elsarticle}
\journal{Journal of Computational Physics}

\newcommand{\V}{\bm{\Phi}}
\newcommand{\La}{\bm{\Lambda}}
\newcommand{\vphi}{\bm{\varphi}}
\newcommand{\absorb}{{\mathbf{A}}}
\newcommand{\absorba}{{\mathbf{A}^\mathrm{a}}}
\newcommand{\scale}{{\mathbf{B}}}
\newcommand{\ones}{\vec{\mathbf{1}}}

\usepackage{stmaryrd}
\usepackage{amsmath}
\usepackage{bm}
\usepackage{amsfonts}
\usepackage{color}
\usepackage{multirow}
\usepackage{mathtools}
\usepackage{lineno,hyperref}
\hypersetup{pdfauthor={Name}}

\usepackage{ifpdf}
\ifpdf
  \DeclareGraphicsExtensions{.pdf,.png,.jpg}
\else
  \DeclareGraphicsExtensions{.eps}
\fi

\usepackage{graphicx}
\graphicspath{{Plots/}{./}}

\begin{document}
\begin{frontmatter}
\title{Absorption kinetics of vacancies by cavities in Aluminum: numerical characterization of sink strengths and first-passage statistics through Krylov subspace projection and eigenvalue deflation}
\author[a]{Savneet Kaur}
\author[a]{Manuel Ath\`enes\corref{cor1}}
\ead{manuel.athenes@cea.fr} 
\cortext[cor1]{Corresponding author}
\author[b]{Jérôme Creuze}
\address[a]{Universit\'e Paris-Saclay, CEA, Service de Recherches de M\'etallurgie Physique, F-91191 Gif-sur-Yvette, France}
\address[b]{SP2M – ICMMO, Universit\'e Paris-Saclay, 91405 Orsay, France}


\begin{abstract}
Modeling the microstructural evolution of metal and alloys, specifically under irradiation, is essential to predict the aging properties of materials. Many models are based on a transition rate matrix describing the jump frequencies of defects and involve a master equation governing the time evolution of a state probability vector. Here, we present non-stochastic numerical techniques to characterize the motion of individual defects migrating over long distances prior to recombining or being absorbed by another defect, resorting to the theory of absorbing Markov chains. These important events are fully determined by their first-passage time distribution to distant locations, no-passage distribution ,and walker fluxes to the sinks. We show that these functions can be efficiently computed using a method combining Krylov subspace projection and eigenvalue deflation. For a model system describing the absorption of a vacancy by a cavity in aluminum, the use of a small Krylov subspace deflated by the unique eigenmode corresponding to the quasi-stationary distribution is sufficient to capture the kinetics of the defect absorption faithfully. This method can be used in kinetic Monte Carlo simulations to perform stochastic non-local moves or in cluster dynamics simulations to compute sink strengths.
\end{abstract}



\end{frontmatter}
\section{Introduction}
Master equations are used extensively to describe many dynamical systems in natural and engineering sciences, in domains ranging from sociology, neuroscience, signal processing, and biochemistry to condensed matter physics. They involve a transition rate matrix and govern the evolution of a high dimensional state probability vector. Introductory examples for processes governed by a master equation are the random walk on an integer line with possible steps in forward, and backward directions, Brownian motion, the path of a diffusing particle until it gets absorbed \cite{buchete_coarse_2008,oppelstrup_first-passage_2009}. The time evolution of condensed matter systems can be simulated directly using one stochastic approach like Langevin dynamics for model systems whose phase space is continuous and Kinetic Monte Carlo (KMC) methods for discrete systems. Any simulation consists of randomly generating typical trajectories in the path space. Some of the physical systems that KMC involves are defect diffusion in alloys, for instance, vacancy motion in general or vacancy/interstitial clustering in ion or neutron-irradiated materials of nuclear reactors~\cite{iracane_jules_2008,carter_microstructural_2001}. While pressure vessels in industrial reactors are usually made of ferritic stainless steels, aluminum alloys are also employed in some experimental reactors. Under irradiation, these materials lose their mechanical properties over time due to the creation of vacancies and interstitials that recombine and form vacancy cavities and interstitial loops. Given its complexity, the master equation associated with the microstructural evolution~\cite{phythian_microstructural_1993,buswell_irradiation-induced_1995} of irradiated materials is often simulated using various KMC methods~\cite{frankcombe_numerical_2009}. 

Several KMC methods are available and are implemented depending on the investigated time and space scales. Atomistic kinetic Monte Carlo (AKMC) methods monitor the positions of all atoms on the lattice~\cite{bortz:1975,ngayam-happy_formation_2012,martinez:2011} or over the space~\cite{nandipati:2012,trochet2020off}. The variant method is commonly known as object kinetic Monte Carlo (OKMC), rather simulate diffusing entities like defect-solute clusters~\cite{domain2020object}. The class of OKMC methods encompasses event Kinetic Monte Carlo (EKMC) methods~\cite{lanore:1974,hoang:2015} and first-passage kinetic Monte Carlo (FPKMC) methods~\cite{opplestrup_first-passage_2006,donev:2010}. The former assumes that all events are possible at any time independently of each other. Simultaneously, the latter introduces spatial protections and enforces the synchronization of the diffusing entities rigorously based on the exact first-passage and no-passage distributions defined hereafter. 

Historically, FPKMC algorithms have been developed to speed up KMC simulations. Indeed, the traditional KMC method may become inefficient when employed to simulate all the hops of defects on a lattice \cite{voter2007introduction} and when the transition rate matrix equation exhibits a broad spectrum of frequencies. The causes of inefficiency may be energetic or entropic in origin. In the former situation, a diffusing defect performs many transitions between a few atomic configurations connected by small energy barriers, typically a vacancy binding to a solute cluster, before escaping elsewhere~\cite{daniels:2020}. These connected configurations form trapping basins. The typical escape time of a defect is much higher than the typical time to cross into the small barriers. Subsequently, the freed vacancy may perform a considerable number of hops in bulk before recombining with another defect or being absorbed: the entropic origin for simulation inefficiency refers to this situation. The statistically exact approach to mitigate the inefficiency of the KMC methods is to draw sequences of events and first-passage times based on the theory of absorbing Markov chains \cite{novotny_monte_1995,opplestrup_first-passage_2006,puchala_energy_2010,donev:2010,athenes_path_2014,nandipati_first-passage_2010,redner2001guide}. Mathematically, the first passage times are the sum of the residence times spent by the walker in connected states before getting absorbed by an artificial or physical sink. The absorbing sink is artificial when it corresponds to the peripheral states of an energetic trap, while a physical absorbing sink usually corresponds to an energetic trapping basin, like solute clusters and dislocations. Furthermore, once the system has reached an absorbing state in absorbing Markov chains, it stays there infinitely. Because the probability of being absorbed tends to one as time tends to infinity, the connected states of the trapping basin are commonly known as transient states. Besides, a defect initially located in any transient state can reach any absorbing state, but not necessarily in one step. The no-passage distribution is the conditional probability distribution of the defect on the transient states knowing that it has not been absorbed yet: the sum of probabilities over the transient states is one. 

Note that master equations associated with absorbing Markov chains also play a crucial role in the formulation and calculation of sink strengths~\cite{carpentier_effect_2017,carpentier_effect_2020,adjanor_complete_2020-1,adjanor_complete_2020}. Sink strengths are important input parameters controlling the reaction rates in cluster dynamics (CD) simulations~\cite{jourdan:2014}. CD equations consist of a high-dimensional set of ordinary differential equations describing the evolution of cluster concentrations. They are obtained through a coarse-graining procedure from the chemical master equation describing the probability vector of cluster populations. Sink strengths are inversely proportional to the mean first passage time and the diffusion coefficient~\cite{malerba_object_2007,carpentier_effect_2017}.

There exist several ways to characterize absorbing Markov chains numerically. The essential goal is to compute the first-passage and no-passage distributions. These two distributions serve to draw the first passage times and moves for a defect to reach the absorbing sink. To achieve these tasks, one implements one of the two following randomization procedures: kinetic path sampling or reverse sampling, based on the factorization~\cite{athenes_path_2014,athenes_elastodiffusion_2019,swinburne_rare_2020} or the eigenvalue decomposition~\cite{novotny_monte_1995,athenes_elastodiffusion_2019} of the absorbing transition rate matrix, respectively. When eigenvalue decomposition is performed~\cite{novotny_monte_1995,puchala_energy_2010,nandipati_first-passage_2010,athenes_elastodiffusion_2019}, the survival probability of the defect on the transient state is computed directly, making it possible to draw the desired first-passage or no-passage times through reverse sampling. The original Markov chain describing the defect evolution is assumed to be reversible. This means that the transport of defects obeys detailed balance; the transition probability flux between any two states is invariant under Markov chain reversal. The transition rate matrix of the Markov chain can therefore be symmetrized using similarity transformation, which ensures the matrix to be symmetric negative semi-definite. The survival probability of the defect before getting absorbed is obtained from the transient evolution operator, a matrix exponential, which is the sum of decaying exponentials. The eigenspectrum of the matrix is real and strictly negative. In this work, the survival probability distribution is estimated using the eigenvalue decomposition method. 

In practice, it is unfortunately difficult to entirely factorize or diagonalize large sparse matrices using dense solvers based on Gaussian elimination, Givens rotations, or Householder reflections due to memory limitations. Krylov subspace projection (KSP) methods are commonly used to obtain solutions for sparse high-dimensional linear systems. The approximations to these solutions are estimated by minimizing the residual over the subspace formed. A well-known KSP method is the conjugate gradient (CG)~\cite{hestenes1952methods}, which is used for solving linear systems involving symmetric and positive definite matrices. For symmetric and possibly indefinite system, iterative method like minimum residual (MINRES) method is rather used~\cite{choi_minres-qlp_2011}. In case of non-symmetric matrices, the biconjugate gradient stabilized (BiCGSTAB) method~\cite{saad2003iterative,amritkar_recycling_2015} that is a generalized CG method, and also a generalized minimum residual (GMRES) method~\cite{amritkar_recycling_2015} are available. KSP methods can also be employed to extract a few pairs of eigenvectors and eigenvalues iteratively, for instance, the Krylov-Schur method~\cite{stewart:2002,arbenz:2016}. 

The eigenvector associated with the smallest eigenvalue is proportional to the quasi-stationary distribution (QSD). It corresponds to the eigenmode exhibiting the slowest decay and limiting the no-passage distribution in the asymptotic time limit~\cite{LeBris:2012,di_gesu_jump_2016}. The QSD is observed to considerably contribute to the first-passage and no-passage distributions in applications~\cite{athenes_elastodiffusion_2019}. It completely characterizes them when trapping is severe and has an energetic origin. However, for purely entropic traps, it is observed that many additional eigenmodes are necessary to capture the early stage absorption kinetics~\cite{athenes_elastodiffusion_2019} correctly. Consequently, the computational cost increases with the number of significant eigenmodes. In this work, we investigate the ability of model order reduction techniques based on Krylov subspace projection and eigenvalue deflation to faithfully characterize the early-stage kinetics at a reduced cost, given an initial probability vector. 

The article is organized as follows. In Section~\ref{mth_fsm}, we introduce the theory of absorbing Markov chains and then formalize the generalized eigenvalue problem. We next describe model order reduction techniques combining Krylov subspace projection and eigenvalue deflation in Section~\ref{model}. In Section~\ref{discuss}, we assess the efficiency of the developed methods by applying them to two problems, a toy model for two-dimensional defect absorption and a realistic model describing the absorption of a distant single vacancy by a cavity in Aluminum~\cite{carpentier_effect_2017}. We discuss the most efficient strategy to compute the probability vector at times shorter than the mean first-passage time depending on the problem. We conclude in Section~\ref{conclusions}.

\section{Theory and methodology} \label{th_mth}
\subsection{Mathematical Formalism}\label{mth_fsm}
A master equation (ME) is the set of ordinary differential equations describing the evolution of the state probability vector $\mathbf{p}(t)$ and writes \cite{redner2001guide}:
\begin{equation}\label{eq:me}
    \dot{\mathbf{p}}^T(t) = \mathbf{p}^T(t) \mathbf{K},
\end{equation}
where $\mathbf{K}$ is the Markov matrix. Its elements $K_{ij}$ are the transition rates from state $i$ to state $j$ and superscript $T$ stands for the transposition. Here, transition rates are assumed to be time independent \cite{athenes_path_2014}. The time evolution operator obtained formally from solutions of the ME can be expressed as
\begin{equation}\label{eq:ev_op}
    \mathbf{P}(t,t^{'})= \exp\left[\int_{t}^{t^{'}} \mathbf{K}dt\right]=\exp((t^{'}-t)\mathbf{K}).
\end{equation}
Matrix element $P_{ij}(t,t^{'})$ represents the probability to find a system in state $j$ at time $t^{'}$ given that it was previously in state $i$ at time $t$. For the models with large number of states, it is not possible to directly compute the solution from the matrix exponential. Nevertheless, this can be done for the absorbing Markov process coinciding with the original process on a subset of $N$ states called transient states and herein indexed from $1$ to $N$. States connected to the transient states are pooled together into a single state labeled as $N+1$ and called absorbing state. The mathematical derivation below follows from Ref.\cite{athenes_elastodiffusion_2019}: the transition rate matrix associated with the absorbing Markov process is cast into the following conventional form: 
\begin{equation}\label{eq:tm_markov}
    \mathbf{K}^\mathrm{a}    =   \left(\begin{array}{c|c}
                - \absorba & \absorba \ones \\ \hline
                   \vec{\mathbf{0}}^T   & 0
              \end{array} \right),
\end{equation}
where $\absorba$ is the $N \times N$ matrix such that $A^\mathrm{a}_{ij}=-K_{ij}$. We define $\vec{\mathbf{1}}=(1,\cdots,1)^T$ as the $N$-dimensional column vector whose components all equal one and similarly $\vec{\mathbf{0}}=(0,\cdots,0)^T$. For the absorbing transition rate matrix $\mathbf{K}^\mathrm{a}$ as well, the sum of the elements of each row is equal to 0. The associated evolution operator is defined as, $\mathbf{P}^\mathrm{a}(t)=\exp[\mathbf{K}^\mathrm{a}t]$. Being a stochastic matrix, it satisfies the following relationship $\forall i \in \llbracket 1,N+1\rrbracket$:
\begin{equation}\label{eq:prob}
    \sum\limits_{j=1}^{N+1} P_{ij}^{\mathrm{a}} (t) = 1.
\end{equation}
Because the absorbing state $N+1$ is taken into account, Eq.\eqref{eq:prob} entails the conservation of total probability. For any vector $\bm{\nu}(t)$ evolving according to the ME
\begin{equation}\label{eq:me_prob}
    \dot{\bm{\nu}}^T(t) = \bm{\nu}^T(t) \mathbf{K}^\mathrm{a},  
\end{equation}
the probability to find the walker in one of the $N+1$ states is also conserved:
\begin{equation}\label{eq:tot_prob}
    \sum\limits_{i=1}^{N+1} \nu_i (t) = 1.
\end{equation}
State labeled $N+1$, the absorbing state, is the only recurrent state of the system, while all the other states are transient. Starting from initial state labeled $i \leq N$, the probability that the system is in transient state $j \leq N$ is given by:
\begin{equation}\label{eq:prob_sys}
    P_{ij}^\mathrm{a}(t) = \mathbf{e}_{i}^{T} \exp({-\absorba t}) \mathbf{e}_{j},
\end{equation}
where $\mathbf{e}_i$ and $\mathbf{e}_j$ are the standard basis vectors. We refer the reader to Ref.~\cite{athenes_elastodiffusion_2019} for details concerning the derivation. 

Here, we assume that the original Markov process is reversible, i.e. obeys the principle of detailed balance which implies that the forward and backward fluxes between any two states, labeled by $i$ and $j$, are equal:
\begin{equation}\label{eq:detail_balance}
    \rho_i K_{ij} = \rho_j K_{ji},
\end{equation}
where $\rho_i$ and $\rho_j$ denote the equilibrium stationary probabilities of the two involved states. When the condition given by Eq.\eqref{eq:detail_balance} is satisfied, the state probability vector $\bm{\rho}$ corresponds to an equilibrium Gibbs-Boltzmann distribution. For the two transient states $i$ and $j$, the condition of detailed balance entails
\begin{equation}\label{eq:flux_rate}
    \sqrt{\rho_i} {A}_{ij}^{\mathrm{a}}/\sqrt{\rho_j} = \sqrt{\rho_j} {A}_{ji}^{\mathrm{a}}/\sqrt{\rho_i}.
\end{equation}
From Eq.\eqref{eq:flux_rate}, a symmetric matrix is defined:
\begin{equation}\label{eq:sym_mat}
    A_{ij}^{B} = s_i s_j \sqrt{\rho_i} {A}_{ij}^{\mathrm{a}}/\sqrt{\rho_i} = A_{ji}^{B}, 
\end{equation}
where $s_i$ and $1/ \sqrt{\rho_i}$ are strictly positive scaling factors. For the sake of concision, we express the two scaling factors using  diagonal definite positive matrices
\begin{equation}\label{eq:sym_mat_s_r}
    \mathbf{S} = \sum_{i=1}^{N}s_i\mathbf{e}_{i}\mathbf{e}_{i}^{T}  
    \quad\mathrm{and}\quad  
    \mathbf{R} = \sum_{i=1}^{N}\frac{1}{\sqrt{\rho_i}}\mathbf{e}_{i}\mathbf{e}_{i}^{T}.
\end{equation}
Matrices $\mathbf{S}$, $\mathbf{R}$ and $\mathbf{B}=\mathbf{S}^2$ commute and are invertible enabling one to define
\begin{equation}\label{eq:mat_A}
     \mathbf{A}^B=\mathbf{S}\mathbf{R}^{-1}\mathbf{A}^{\mathrm{a}}\mathbf{R}\mathbf{S}=(\mathbf{SR})^{-1}\mathbf{BA^{\mathrm{a}}}(\mathbf{RS}).
\end{equation}
Matrix $\mathbf{B}$ is used as a preconditioner, i.e. it aims at decreasing the condition number of $\mathbf{A}^B$. Matrix $\mathbf{SR}$ serves to define a similarity transformation between $\absorb^B$ and $\scale \absorb^{\mathrm{a}}$ and to formulate a generalized symmetric eigenvalue problem. A similarity transformation preserving spectral properties, the eigenvalues of $\mathbf{BA^{\mathrm{a}}}$ are real and definite positive. This is true in particular for $\absorb^{\mathrm{a}}$ when $\scale$ is set to the identity $\mathbf{I}$.

The spectrum of $\mathbf{A}^{\mathrm{a}}$ can be obtained by solving the following generalized eigenvalue problem:
\begin{equation}\label{eq:ev}
    \mathbf{A}^B \vphi_k = \mathbf{B} \vphi_k \lambda_k,
\end{equation}
where $\vphi_k$ denotes the $k$th eigenvector associated with $\lambda_k$, the $k$th eigenvalue. We assume that eigenvalues are sorted in ascending order: $0 <\lambda_1 \leq \lambda_2 \leq \cdots \lambda_N$ and that eigenvectors are normalized. Hence, $\mathbf{\Phi}=(\bm{\varphi}_1,\bm{\varphi}_2,...,\bm{\varphi}_N)$ is an orthonormal basis of eigenvectors. From  Eq.\eqref{eq:ev}, we obtain
\begin{equation}\label{eq:ev_spec}
    \mathbf{A}^B \V = \mathbf{B} \V \La,
\end{equation}
where $\mathbf{\Lambda}$ is a diagonal matrix with eigenvalues as its diagonal elements ($\varLambda_{ii}=\lambda_i$). Multiplying both the left and right sides by $\mathbf{\Phi}^T$ results in,
\begin{equation}\label{eq:orthogonality}
    \V^T\mathbf{A}^B\V = \mathbf{\Lambda}, \qquad \mathbf{A}^B = \mathbf{B}\V\La \V^T \mathbf{B},
\end{equation}
where the $\scale$ orthogonality of \bm{$\Phi$} has been taken into account
\begin{equation}\label{eq:ortho_consequence}
    \V^T\mathbf{B}\V = \mathbf{I}, \qquad \V\V^T = \mathbf{B}^{-1}.
\end{equation}
Now, the absorbing transition rate matrix can be obtained from Eq.\eqref{eq:mat_A} and Eq.\eqref{eq:orthogonality} as
\begin{equation}\label{eq:tra_mar_mat_ev_spec}
    \absorba  = \mathbf{RS} \V \La \V^T \mathbf{SR}^{-1}.
\end{equation}
From the matrix decomposition \eqref{eq:tra_mar_mat_ev_spec}, the exponential of the transition rate matrix is expressed as
\begin{equation}\label{eq:exp_tra}
    \exp\left[ -\absorba t \right] = \mathbf{RS} \V\exp\left[ - \mathbf{\Lambda} t\right] \V^T \mathbf{SR}^{-1}.
\end{equation}
We know introduce $\mathbf{s}$ and $\mathbf{r}$, the scaling and re-scaling vectors consisting of diagonal elements of matrices $\mathbf{S}$ and $\mathbf{R}$, respectively. We have $\mathbf{r}=\mathbf{R}\mathbf{\vec{1}}$ and $\mathbf{s}=\mathbf{S}\mathbf{\vec{1}}$ and also define two additional sets of re-scaled basis vectors $\mathbf{g_i}=\mathbf{e}_i\odot\mathbf{s}\odot \mathbf{r}$ and $\mathbf{d_j}=\mathbf{e}_j\odot\mathbf{s}\oslash \mathbf{r}$, where $\odot$ and $\oslash$ denote component-wise multiplication and division. 

Plugging the scalar products $g_i^h=\bm{\varphi}_h^T\mathbf{g}_i$ and $d_j^h=\bm{\varphi}_h^T\mathbf{d}_j$ into the evolution operator, Eq.\eqref{eq:prob_sys} gives
\begin{equation}\label{eq:prob_exp_tra}
    P_{ij}^\mathrm{a}(t)=\sum_{h=1}^N g_i^h d_j^h \exp(-\lambda_ht),
\end{equation}
where $(i,j) \leq N$. The survival probability of a system that has evolved from the initial state $i$ at time $t=0$ is the probability that the system has not been absorbed yet, or equivalently, is still located in one of the transient state. From Eq.\eqref{eq:prob_exp_tra}, it is easy to deduce 
\begin{equation}\label{eq:survival_probability}
    p^{\mathrm{s}}_i(t) =\sum_{j=1}^N P_{ij}^\mathrm{a}(t)=\sum_{j=1}^N \sum_{h=1}^N g_i^h d_j^h \exp(-\lambda_ht) = \sum_{h=1}^N \alpha_i^h \exp(-\lambda_ht),
\end{equation}
where $\alpha_i^h = g_i^h \sum_{j=1}^N d_j^h$. In Section~\ref{discuss}, the survival probability is simply denoted by $S(t)=\bm{\pi}^T\mathbf{p}^{\mathrm{s}}(t)$, omitting the dependence on the initial state $\bm{\pi}$. 
The mean first-passage time (MFPT) vector, $\bm{\tau}^{(N)} = \mathbf{RSA}^{-1}\mathbf{SR}^{-1}\ones$, corresponds to the solution of a linear system of $N$ equations. Besides, the MFPT from initial state $i$ exhibits a simple expression in the eigen basis: 
\begin{align}\label{def:mfpt}
    \tau_i^{(N)} = \sum_{h=1}^N \alpha_i^h/\lambda_h. 
\end{align}

\subsection{Quasi-stationary distribution and residence times} \label{qsd_rt}
The no-passage probability distribution is defined by $P_{ij}^\mathrm{np}(t)= P_{ij}^\mathrm{a}(t)/p^{\mathrm{s}}_i(t)$ for $i,j \in \llbracket 1 , N\rrbracket$. Likewise, the QSD is defined as the probability vector $q_j$ that behaves asymptotically in time as the no-passage distributions. This is independent of initial state $i$
\begin{equation}\label{def:qsd}
    q_j = \lim_{t \rightarrow \infty} P_{ij}^\mathrm{np}(t) = \frac{d^1_j}{\sum_{h=1}^N d^1_h} 
\end{equation}
The MFPT is obtained by integrating over time:
\begin{equation}\label{def:mfpt_1}
      \tau_i^{(N)} = \int_0^{\infty} t \frac{d}{dt}p_i^\mathrm{a}(t) dt = \mathbf{e}_i^T \mathbf{A}^{-1}\ones,
\end{equation}
where the quantity $p^\mathrm{a}_{i}(t) = 1-p^\mathrm{s}_{i}(t)$ corresponds to the probability that the walker has been absorbed at $t$. The absorption probability $p_i^\mathrm{a}$ increases from 0 at $t=0$ to 1 at $t=\infty$. The time derivative of $p^\mathrm{a}_i(t)$ corresponds to the probability distribution of first-passage times.

\subsection{Model Order Reduction and Iterative Methods}\label{model}
\subsubsection{Eigenvector Subspace Model Projection} \label{esmp}
As mentioned in the introduction, it is not possible to compute the entire eigenspectrum of huge sparse matrices using standard dense solvers based on Givens rotation or Householder reflections for memory limitations. Instead, sparse and iterative eigenvalue solvers \cite{slepc-users-manual,petsc-user-ref,petsc-web-page} are to be implemented to extract a limited portion of the eigenspectrum and to approximate the evolution operator by its projection on the reduced eigenbasis \cite{athenes_elastodiffusion_2019}. In practice, the basis encompasses the $k$ eigenvectors with smallest eigenvalues. By doing so, a reduction of the model is performed and the symmetric transition rate matrix $\mathbf{A}^a$ in Eq.\eqref{eq:tra_mar_mat_ev_spec} is approximated by 
\begin{equation}\label{eq:esmp_ev_spec}
    \mathbf{A}^{(k)} = \mathbf{SR} \left[\sum_{h=0}^k \vphi_h\vphi^T_h\lambda_h \right] \mathbf{SR}^{-1}, 
\end{equation}
and the evolution operator is approximated by
\begin{equation}\label{eq:esmp_evo_op}
    P_{ij}^{(k)}(t)=\sum_{h=1}^k g_i^h d_j^h \exp(-\lambda_ht). 
\end{equation}
In the following, we will refer to this approach as the Eigenvector Subspace Model Projection (ESMP) method. Assessing the convergence as a function of the eigenvalue number $k$ can be done by calculating the survival probability on the reduced eigenvector space and comparing the result with the one obtained using a standard solver computing the matrix exponential function with a built-in convergence criteria~\cite{eiermann_restarted_2006}. An additional verification can be performed by comparing the truncated MFPT 
\begin{align}
 \tau_i^{(k)} = \sum_{h=1}^k \alpha_i^h \lambda_h
\end{align}
to the exact one, i.e., to the value $\tau_i^{(N)}$ defined in Eq.~\eqref{def:mfpt} and computed using two well-established and robust methods: conjugate gradient and multi-frontal Cholesky~\cite{amestoy_fully_2001} algorithms. 

In the following subsections, we present additional model reduction approaches based on Krylov subspace projection methods: refer to~\cite[Ch. 7]{saad2003iterative} for a textbook. Notice that $\absorb \in \mathbb{R}^{N\times N}$ will stand for $\absorb^I$, i.e. the scaling matrix $\mathbf{B}$ defined in previous section for the sake of generality will always be equal to the identity to restrict our investigation.   

\subsubsection{Krylov Subspaces} \label{mfn_ks}
Evaluating matrix functions via eigenvalue decomposition becomes extremely costly for large sparse matrices whenever many eigenpairs are needed. A solution to this problem is to restrict the computation to the product of a vector a matrix function of  $\absorb \in \mathbb{R}^{N \times N}$ an a vector $\mathbf{b} \in \mathbb{R}^N$: 
\begin{equation}\label{eq:ksmp_y}
    \mathbf{x} = f(\absorb)\mathbf{b}.
\end{equation}
This task is efficiently accomplished using a Krylov subspace projection (KSP) method \cite{ipsen_carl}. In particular, KSP methods are among the most efficient algorithms for estimating the solution of huge sparse linear systems. As examples of Krylov subspace solvers, one mentioned in the introduction, the conjugate gradient and minimal residual methods aim at computing the quantity
\begin{equation}\label{eq:ksmp_ax}
   \mathbf{x}= \absorb^{-1}\mathbf{b}, 
\end{equation}
which amounts to setting $\mathbf{f}(\absorb)=\absorb^{-1}$~\cite[Ch. 9]{saad2003iterative}. The Krylov subspace is characterized by its dimension $\ell$, matrix $\absorb$, and vector $\mathbf{b}$ and is denoted by $\mathcal{K}_\ell(\absorb,\mathbf{b})$. Its construction proceeds as follows: vector $\mathbf{b}$ is left multiplied by $\absorb$, which results in a new vector $\absorb \mathbf{b}$. The new vector is multiplied again with matrix $\absorb $ to find $\absorb^2\mathbf{b}$, and this goes on $\ell-1$ times, vector $\mathbf{b}$ being included in the subspace: 
\begin{equation}\label{eq:krylov_subspace}
    \mathcal{K}_\ell(\absorb,\mathbf{b})=\mathrm{span} \big\{\mathbf{b},\absorb\mathbf{b},\absorb^2\mathbf{b},...,\absorb^{\ell-1}\mathbf{b}\big\}.
\end{equation}
Gram-Schmidt algorithm is also used on the fly to construct an orthogonal basis of $\mathcal{K}_\ell (\absorb,\mathbf{b})$ whose dimension is always equal to $\ell$. If it happens that $\absorb^l \mathbb{b} \in \mathcal{K}_{l} (\absorb,\mathbf{b})$, the subspace construction is resumed and $\ell$ is set to $l$. The basis of $\mathcal{K}_\ell (\absorb,\mathbf{b})$ is denoted by $\mathbf{V}_\ell$. 

\subsubsection{Krylov Subspace Model Projection} \label{ksmp}
Using the definition of Krylov subspace from Section~\ref{mfn_ks}, the standard scheme involving matrix-vector multiplications is recalled. In practice, the problem of computing the exponential of a huge sparse matrix $\absorb \in \mathbb{R}^{N,N}$ is reduced to the one of computing the exponential of a small matrix $\mathbf{T}_\ell$ of dimension $\ell$ 
\begin{equation}\label{eq:func_A}
    \absorb \mathbf{V}_\ell = \mathbf{V}_{\ell} \mathbf{T}_\ell + T_{\ell+1,\ell} \mathbf{v}_{\ell+1} \widehat{\mathbf{e}}_\ell^T,
\end{equation} 
where $\mathbf{V}_\ell = \big[ \mathbf{v}_1,\mathbf{v}_2,\dots,\mathbf{v}_{\ell} \big]$ consists of $\ell$ orthonormal column vectors and $\mathbf{T}_\ell \in \mathbb{R}^{\ell,\ell}$ is a symmetric tri-diagonal matrix. Since the vectors from $\mathbf{V}_\ell$ basis are orthonormal, we have $\mathbf{V}_{\ell}^T \mathbf{V}_{\ell} = \mathbf{I}_{\ell}$ where $\mathbf{I}_{\ell} \in \mathbb{R}^{\ell,\ell}$ is the identity matrix. 
$\mathbf{T}_\ell$ corresponds to the projection of $\absorb$ onto  $\mathbf{V}_\ell$, and $\widehat{\mathbf{e}}_\ell^T$ denotes the $\ell$th unit coordinate vector of $ \mathbb{R}^{\ell}$. The reduced tridiagonal matrix is obtained by left-multiplying both sides of  Eq.\eqref{eq:func_A} by $\mathbf{V}_{\ell}^T$: 
\begin{equation}\label{eq:projection}
    \mathbf{V}_{\ell}^T \absorb \mathbf{V}_\ell = \mathbf{V}_{\ell}^T \mathbf{V}_{\ell} \mathbf{T}_\ell =\mathbf{T}_\ell , 
\end{equation}
while the Arnoldi approximation of $\absorb$ is $\mathbf{V}_\ell \mathbf{T}_\ell \mathbf{V}_\ell^T$. This last matrix is then used to approximate $f(\absorb)\mathbf{b}$ using the following vector  
\begin{equation}\label{eq:function_reduced_matrix}
   \mathbf{f}_\ell = f \left( \mathbf{V}_\ell \mathbf{T}_\ell \mathbf{V}^T_\ell \right) \mathbf{b} = \beta \mathbf{V}_\ell f(\mathbf{T}_\ell) \widehat{\mathbf{e}}_1,
\end{equation}
where $\widehat{\mathbf{e}}_1$ stands for the first unit coordinate vector of $\mathbb{R}^\ell$, $\beta$ for $\|\mathbf{b}\|$, the Euclidean norm of $\mathbf{b}$, and where we plugged the relation $\mathbf{b} = \beta\mathbf{V}_{\ell}\widehat{\mathbf{e}}_1$. 

This approach is another example of model order reduction. For a sparse matrix, the complexity is $\mathcal{O}(N\ell)$ for storage and $\mathcal{O}(N^2\ell)$ for the $\ell$ steps of iteration \cite{antoulas_approximation_2005,higham_computing_2010}. Krylov subspaces $\mathcal{K}_\ell(\absorb,\mathbf{b})$ of low dimensions are used in practice, hence this approach is considerably faster than exact full diagonalization techniques that exhibit $\mathcal{O}(N^3)$ complexity in number of operations and $\mathcal{O}(N^2)$ in storage.

\subsubsection{Eigenvector and Krylov Subspace Model Projection}\label{eksmp}
In this section, we describe a third approach, referred to as eigenvector and Krylov subspace projection method (EKSMP), consisting in projecting the model both on eigenvector and Krylov subspaces. It is based on the standard deflation technique of linear algebra \cite{coulaud_deflation_2013,article,gaul_framework_2013}. 
In deflation, the approximation subspace is divided into two complementary subspaces, so that the two parts of the solution are easier to find using an exact method in the first subspace and an iterative method in the second deflated subspace. 

The deflation approach is usually implemented for solving linear systems but we consider it here for evaluating the application of any matrix function on a vector, i.e. $f(\absorb)\mathbf{b}$. As previously, function $f$ will be either the inverse or scaled exponential functions. The main objective of this scheme is to obtain a deflated matrix $\absorb^\perp$ whose condition number will be smaller than that of $\absorb$. This is done by discarding the contribution of a few smallest eigenvalues from the system, and focusing on a deflated matrix. Hence, the first subspace is generated by the $k$ eigenvectors of $\mathbf{A}$ associated with the lowest eigenvalues as in Section~\ref{esmp} and is denoted by $\mathcal{E}_k(\mathbf{A})$. The deflated subspace is the orthogonal component of the eigenvector subspace and is denoted by $\mathcal{E}_k(\mathbf{A})^\perp$. The Krylov subspace is then constructed in the deflated space. The goal is to accelerate the convergence of the projected dynamics towards the exact solution as the dimension $\ell$ of the Krylov subspace increases. 

An orthogonal basis of $\mathcal{E}_k(\mathbf{A})$ writes $\V_k = \big[ \bm{\varphi}_1,\bm{\varphi}_2,\dots,\bm{\varphi}_k \big]$, entailing that $\mathbf{P} = \V_k\V_k^T$ and $\mathbf{I}-\mathbf{P}$ are the orthogonal projection operators on $\mathcal{E}_k(\mathbf{A})$ and $\mathcal{E}_k(\mathbf{A})^\perp$. A general property of projection operators is that they are involution, i.e. $\mathbf{P}^n=\mathbf{P}$. A particular property of $\mathbf{P}$ and  $\mathbf{I}-\mathbf{P}$ is that they commute with $\mathbf{A}$, as a result of the spectral theorem. Hence, any power of $\mathbf{A}$ can be decomposed as 
\begin{align}\label{eq:eksmp_decomposition}
    \absorb^n = (\mathbf{PA})^n\mathbf{P} + (\mathbf{(I-P)A})^n(\mathbf{I-P}) =\left(\absorb^{\parallel}\right)^n\mathbf{P}+\left(\absorb^{\perp }\right)^n(\mathbf{I}-\mathbf{P}),
\end{align}
where $\absorb^{\parallel}=\mathbf{PA}$, $\absorb^{\perp}=\absorb-\absorb^{\parallel}$. Consequently, the desired quantity can be decomposed as the sum of the following two terms:
\begin{align}\label{eq:eksmp_decomposition_function}
    f\mathbf{(A)b} =  f \left(\absorb^{\parallel}\right) \mathbf{b}^{\parallel}+  f\mathbf{\left(\absorb^{\perp}\right) {b}^{\perp}}, 
\end{align}
where $\mathbf{b}^{\parallel}$ and $\mathbf{b}^{\perp}$ stands for $\mathbf{Pb}$ and $\mathbf{(I-P)b}$, respectively. Note that the Krylov subspace in the deflated space can be simply generated from the projected initial vector $\mathbf{(I-P)b}$, which is formally stated by 
\begin{align}\label{eq:krylov_deflated_subspace}
    \mathcal{K}_\ell \left( \absorb^{\perp} , \mathbf{b}^{\perp} \right) = \mathcal{K}_\ell(\mathbf{A}, \mathbf{b}-\mathbf{Pb}). 
\end{align}
As a result, the approximation subspace is the sum of two subspaces:
\begin{align}\label{eq:eksmp_approx}
    \mathcal{G}_{k,\ell} (\mathbf{A},\mathbf{b}) = \mathcal{E}_k(\mathbf{A}) \oplus \mathcal{K}_\ell(\mathbf{A}, \mathbf{b}-\mathbf{Pb}),
\end{align}
which are orthogonal to each other. Note that ESMP method considers the first subspace only [Section~\ref{esmp}], while KSMP method the second one only [Section~\ref{ksmp}]. The second term in Eq.\eqref{eq:eksmp_decomposition_function} is projected onto the orthogonal Krylov subspace $\mathcal{K}_\ell \left( \absorb^{\perp} , \mathbf{b}^{\perp} \right)$ and evaluated using full eigenvalue decomposition. The matrix that must be diagonalized is
\begin{equation}
    \mathbf{T}_\ell^\perp=\big(\mathbf{V}_\ell^{\perp}\big)^T \absorb\mathbf{V}_\ell^{\perp}, 
\end{equation}
where $\mathbf{V}_\ell^{\perp}$ is the standard orthogonal basis of $\mathcal{K}_\ell \left( \absorb^{\perp} , \mathbf{b}^{\perp} \right)$. The accuracy of the model order reduction method can be verified by checking the convergence of survival probability distribution and estimation of MFPTs as a function of $\ell$ given $k$. The survival probability of a system that has evolved from initial probability vector $\bm{\pi}$ at time $t=0$ evaluated using EKSMP method is
\begin{equation}\label{eq:eksmp_sp}
    S_{k,\ell}(t) = \sum_{i=1}^N \sum_{h=1}^k \pi_i\alpha_i^h \exp(-\lambda_ht) + \sum_{i=1}^N \sum_{h=1}^\ell \pi_i \widehat{\alpha}_i^{h} \exp(-\lambda^\perp_ht) 
\end{equation}
where $\lambda^\perp_h$ is the $h$th eigenvavalue of $\mathbf{T}_\ell$ and the weighting coefficient $\widehat{\alpha}^h_i$ involves the corresponding $h$th eigenvector $\widehat{\bm{\varphi}}_h$ in the Krylov subspace. After projecting with operator $\widehat{\bm{\varphi}}_h^T \mathbf{V}_\ell^T $, we obtain $\widehat{\alpha}_i^h = \widehat{g}^h_i \widehat{d}^h$ with $\hat{g}_i^h = \widehat{\bm{\varphi}}_h^T \mathbf{V}_\ell^T \mathbf{s \odot r \odot e}_i$ and $\widehat{d}^h = \widehat{\bm{\varphi}}_h^T \mathbf{V}_k^T \mathbf{s\oslash r}$. 
Note that assuming exact arithmetic, the survival probability $S(t) = \bm{\pi}^T\mathbf{p}^\mathrm{s}(t)$ defined from  Eq.\eqref{eq:survival_probability} is equal to $S_{N-\ell,\ell}(t)$ in Eq.\eqref{eq:eksmp_sp}, $\forall \ell\in \llbracket 0 , N \rrbracket$, since the approximation space $\mathcal{G}_{N-\ell,\ell}(\mathbf{A},\bm{\pi})$ spans the entire phase space.

To later monitor the convergence of the methods as a function of $k$ and $\ell$, we will first inspect the estimated survival probability at $t=0$ and additionally evaluate the following reduced MFPT: 
\begin{equation}\label{eq:r.2}
    T_{k,\ell}=\frac{\sum_{i=1}^N \pi_i\left[ \sum_{h=1}^k \alpha^h_i/\lambda_h + \sum_{h=1}^{\ell} \widehat{\alpha}^h_i /\lambda^\perp_h\right]} {\sum_{i=1}^N \pi_i\tau_i^{N}},
\end{equation}
where the denominator in Eq.\eqref{eq:r.2} corresponds to the MFPT from initial distribution $\bm{\pi}$. The MFPT's have been computed using two distinct linear solvers (sparse Cholesky and CG) to check that matching results are obtained. 

\subsection{Implementation}
The methods described above have been coded in PETSC/SLEPc environments resorting to the matrix function (MFN) object \cite{slepc-users-manual,slepc-toms,str-6,petsc-efficient,petsc-user-ref,petsc-web-page}. MFN object also provides the restarted Krylov subspace projection method (R-KSP)~\cite{eiermann_restarted_2006} to compute the product of common matrix functions and a vector. 
R-KSP was used to compute reference values to check the correctness of the faster methods described above. R-KSP is a robust and well-established KSP solver with built-in convergence criteria \cite{eiermann_restarted_2006,str-6} wherein the Krylov basis is restarted until a convergence criterion is fulfilled. We refer the reader to \ref{rksp} for details. In our case, the application of vector $\mathbf{b}$ on the matrix exponential is computed for a predetermined set of times $\Big\{ t_n \Big\}_{0 \leq n \leq L}$: 
\begin{equation}\label{eq:rksp_approx_fin}
    \mathbf{x} = \exp({-\mathbf{A}t_n})\mathbf{b}. 
\end{equation}
The operation must be repeated at each considered time. For this purpose, R-KSP method is much more expensive than EKSMP method, because it is not able to provide the entire first-passage law at once.

\section{Results and Discussions}\label{discuss}
In this section, we first discuss the scalability and efficiency of the four different solvers used to extract eigenvalues. We set up a simple absorption model in two dimensions in Section~\ref{ce}. Next, we illustrate the three computational methods discussed in Section~\ref{model} by applying them to a realistic problem, the absorption of a single vacancy in a cavity in aluminum in Section~\ref{sp_results}. The model describing thermally activated jumps of aluminum atoms into a next nearest-neighbor vacancy is detailed in Ref.~\cite{athenes_elastodiffusion_2019}. It accounts in particular for the dipole-dipole elastic interactions between the vacancy and the cavity~\cite{carpentier_effect_2017}. We characterize the absorption kinetics of the vacancy around the cavity and quantify the effect of the elastic interactions on the vacancy flux towards the cavity in Sec.~\ref{rtv_results}, and on sink strengths (Sec.~\ref{ssf_results}). 

\subsection{Efficiency and scalability of eigensolvers}\label{ce}
The simple absorption model describes the motion of a defect on a periodically replicated square lattice of size $\mathrm{L}$ and coordination number $Z=4$. The defect hops from any site to any of its four nearest neighboring sites with reduced frequency of 1. The number of transient states is $N=\mathrm{L}^2$. In addition, the defect can reach the absorbing sink from a singularized site with an absorbing frequency equal to $10^{-2}$. In this particular application, the considered transition rate matrix $\mathbf{A}$ is thus a modified Laplacian matrix: diagonal elements are equal to 4, but one element is set to value $4.01$, while the $4N$ off-diagonal elements corresponding to transitions are all equal to -1. The matrix $\mathbf{A}$ is thus symmetric positive definite by construction. We herein evaluate the cost of extracting the linear system using the various sparse iterative solvers from SLEPC library~\cite{slepc-users-manual} in PETSC environment~\cite{petsc-user-ref} and with varying the number of cores in the computations. This enables one to deduce the speedup and efficiency resulting from implementing the solvers on a parallel computer architecture. 

\begin{figure}[!ht]
    \centering
     \includegraphics[width=0.7\textwidth]{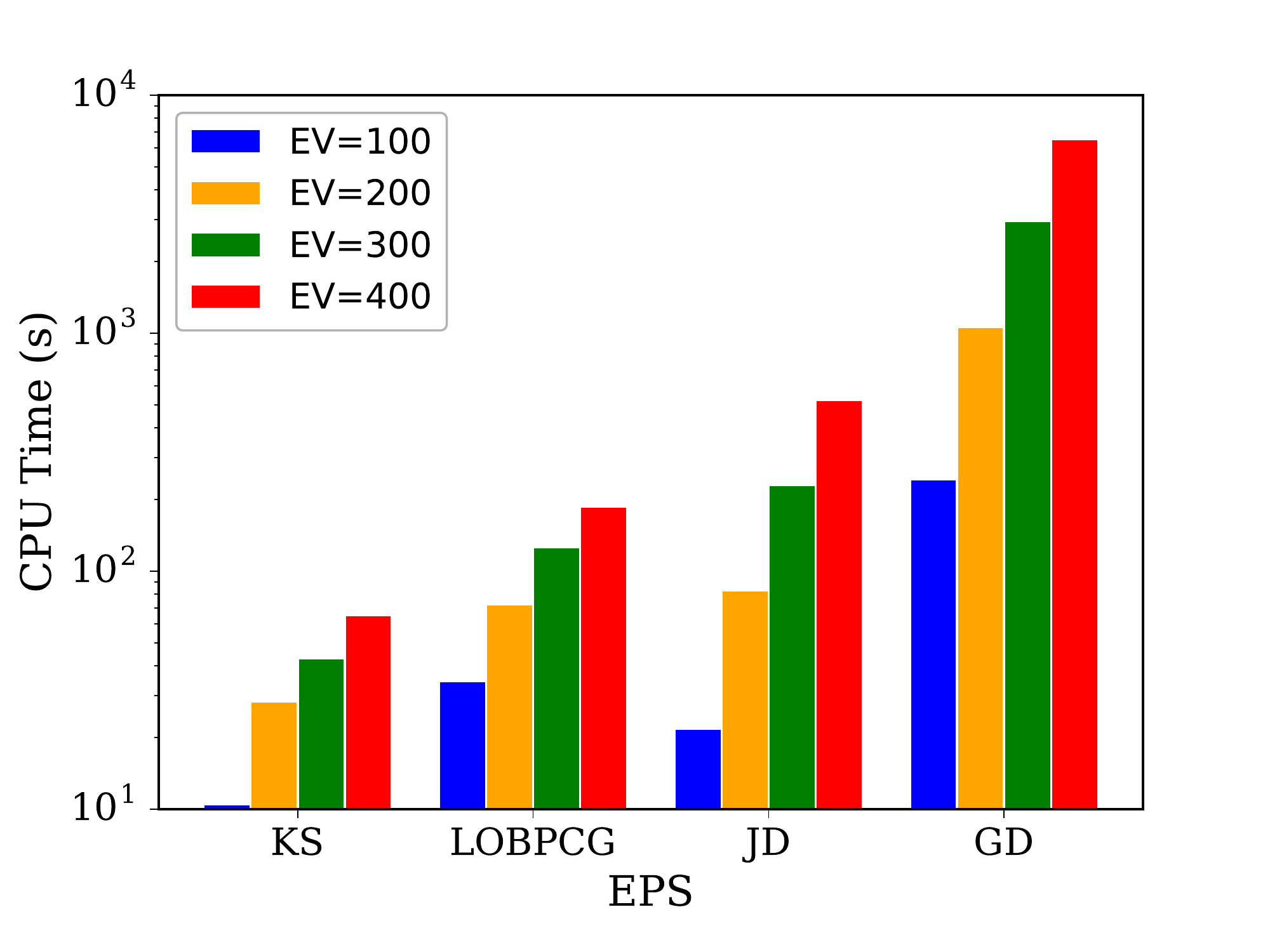}
    \caption{Comparison of CPU cost for four different iterative solvers (EPS) based on extracting the indicated number of requested eigenvalues (EV).}
   \label{fig:diff_solver}
\end{figure}

We use an in-house cluster composed of 36 processing nodes, each node consisting of a dual-processor Intel(R) Xeon(R) Gold 6132 CPU running at 2.60GHz with $2\times14$ cores. As for the sparse iterative solvers, we tested methods like Krylov-Schur (KS), locally optimal block preconditioned conjugate gradient (LOBPCG), generalized Davidson (GD), and Jacobi Davidson (JD). In this first series of computations, the lattice size L is set to the value of $10^2$, so the number of transient states is $N=10^4$. The eigenvalues to be considered sufficiently accurate and converged, the tolerance value to $10^{-10}$. The maximum number of iterations was set to $6 \times 10^4$. KS and LOBPCG solvers are implemented without preconditioning, while block Jacobi preconditioning is used by GD solver and also JD solver within its built-in MINRES routine. Figure~\ref{fig:diff_solver} displays the CPU time required to compute the indicated number of eigenvalues using the four iterative solvers. We observe that the Krylov-Schur solver takes less time to extract the eigenvalues than LOBPCG, JD, and GD solvers do.

The performance of KS, LOBPCG, JD, and GD solvers are all expected to be positively impacted through preconditioning.  We study the impact of preconditioning resorting to the very efficient sparse Cholesky solver from MUMPS, and report results for KS solver only in the following. Unfortunately, MUMPS preconditioner could not be enabled within LOBPCG in SLEPC, or sub-performed  with JD and GD solvers.

\begin{figure}[!h]
    \centering
    \includegraphics[width=0.7\textwidth]{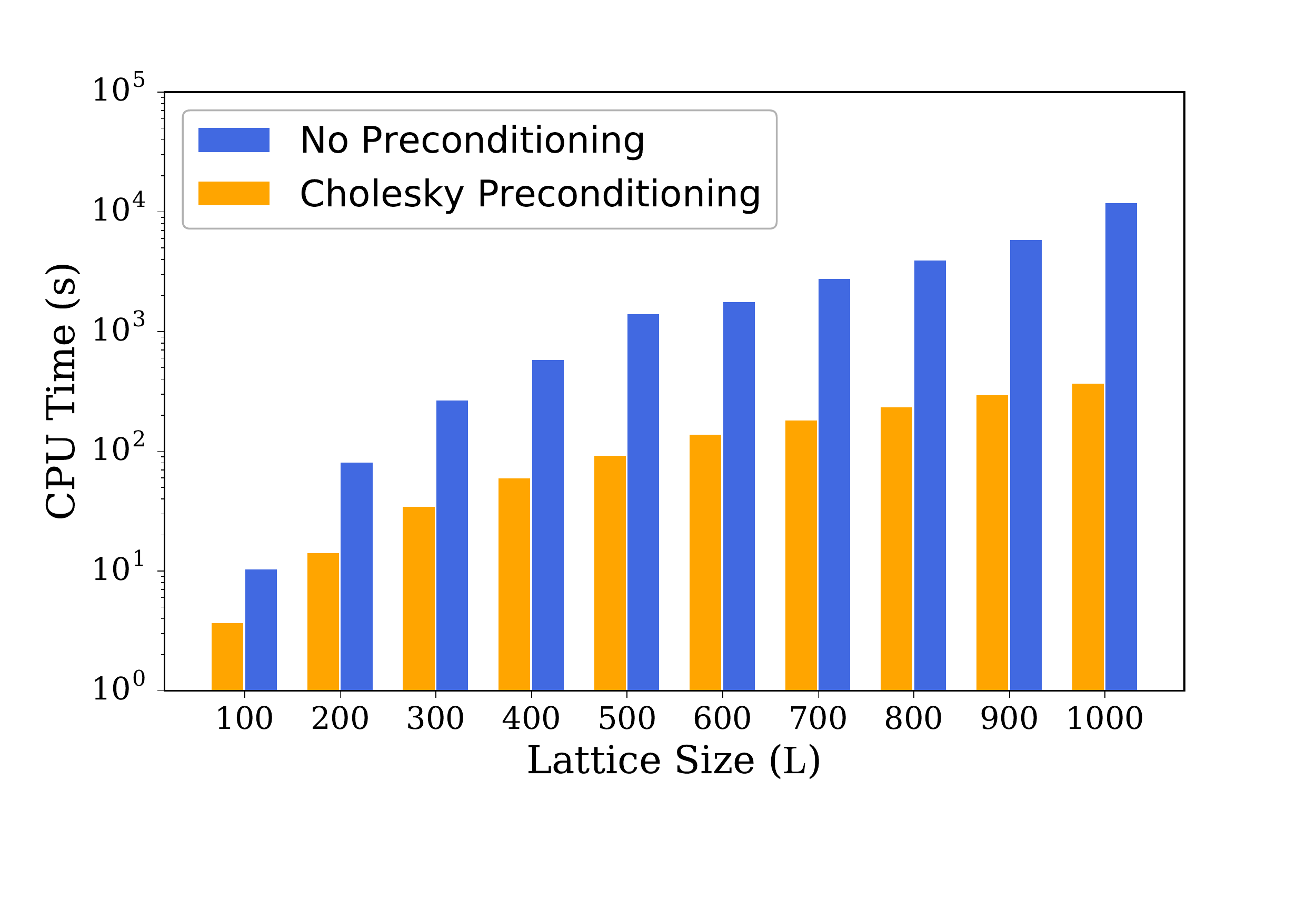} 
    \caption{CPU times for extracting 100 eigenvalues using Krylov-Schur method without preconditioning and with Cholesky preconditioning as a function of the cell size ($\mathrm{L}$)}
   \label{fig:preconditioner_krylovschur}
\end{figure}

For extracting the lowest eigenvalues using the preconditioned KS solver, the inverse spectral transform is enabled and the linear system arising at each reversed iteration is solved using the Cholesky method of MUMPS. Cholesky factorization~\cite{haddad_cholesky_2009,floudas2008encyclopedia} is done once using the MUMPS package~\cite{amestoy_fully_2001}. We check the performance of Krylov-Schur with sparse Cholesky preconditioning by extracting 100 eigenvalues with varying the lattice size $\mathrm{L}$. The tolerance parameter to extract converged eigenvalues is set to be $10^{-10}$. A linear speedup can be seen in Fig.~\ref{fig:preconditioner_krylovschur} for the performance of Krylov-Schur if implemented with Cholesky preconditioning.

\begin{figure}[!ht]
    \centering
    \includegraphics[width=0.6\textwidth]{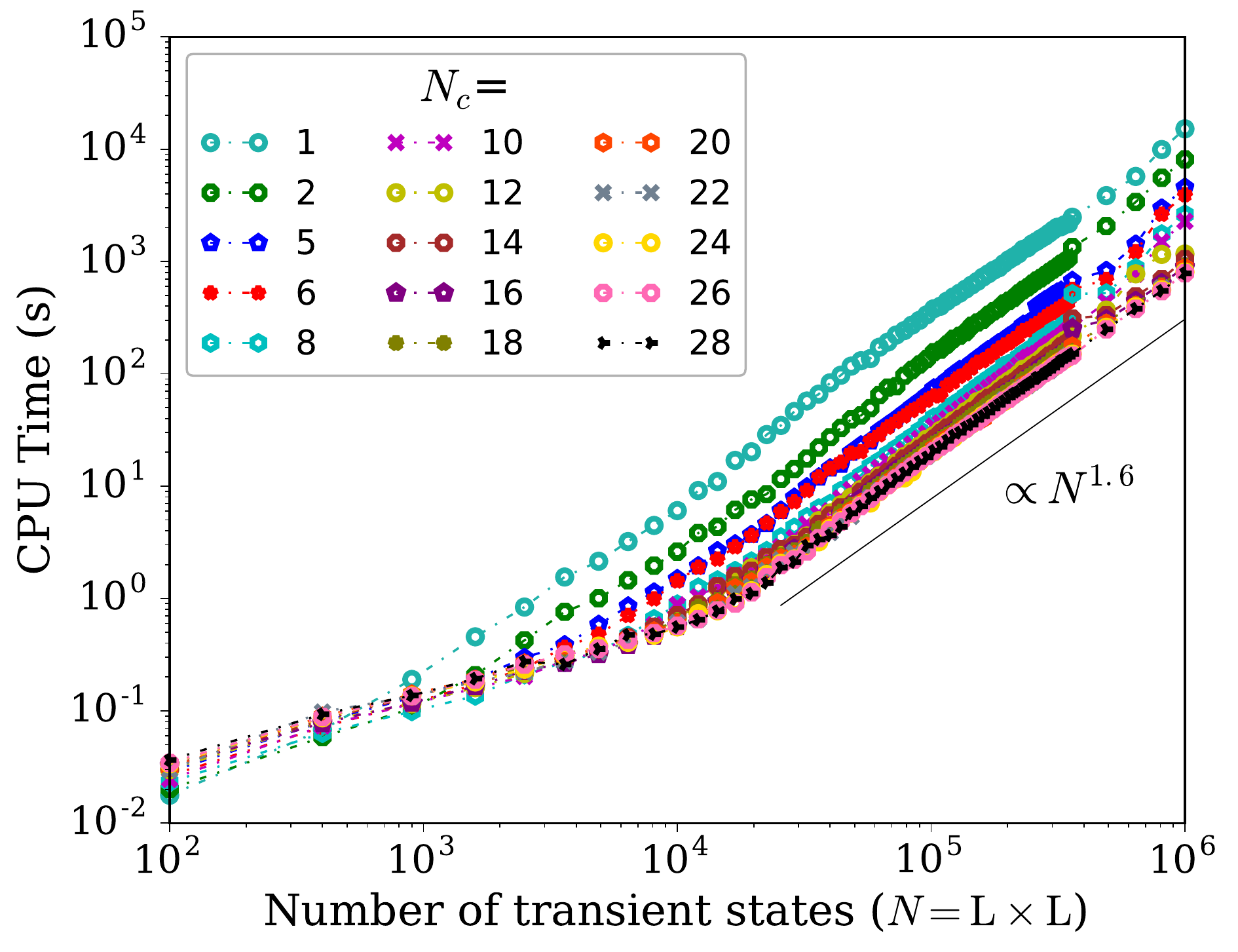} 
    \caption{Evaluated cost for extracting 100 eigenvalues using Krylov-Schur method in sequential and parallel computations as a function of number of transient states $N = \mathrm{L} \times \mathrm{L}$ with varying the number of cores ($N_c$).}
   \label{fig:time_profile}
\end{figure}

As the number of transient states $N$ increases, it gets computationally more expensive to perform a sequential simulation to extract the eigenvalues. To check the scalability of the Krylov-Schur method, we performed parallel computations with 28 cores to extract the first smallest 100 eigenvectors and corresponding eigenvalues for the transient states $N$ increasing from $10^2$ to $10^6$. In Fig.~\ref{fig:time_profile}, we observe a linear decrement in the cost of extracting eigenvalues as the number of cores and the number of transient states $N$ are increased. It can be observed the complexity of the Krylov-Scur method is between square and linear.

\begin{figure}[!ht]
    \centering
    \includegraphics[width=0.5\textwidth]{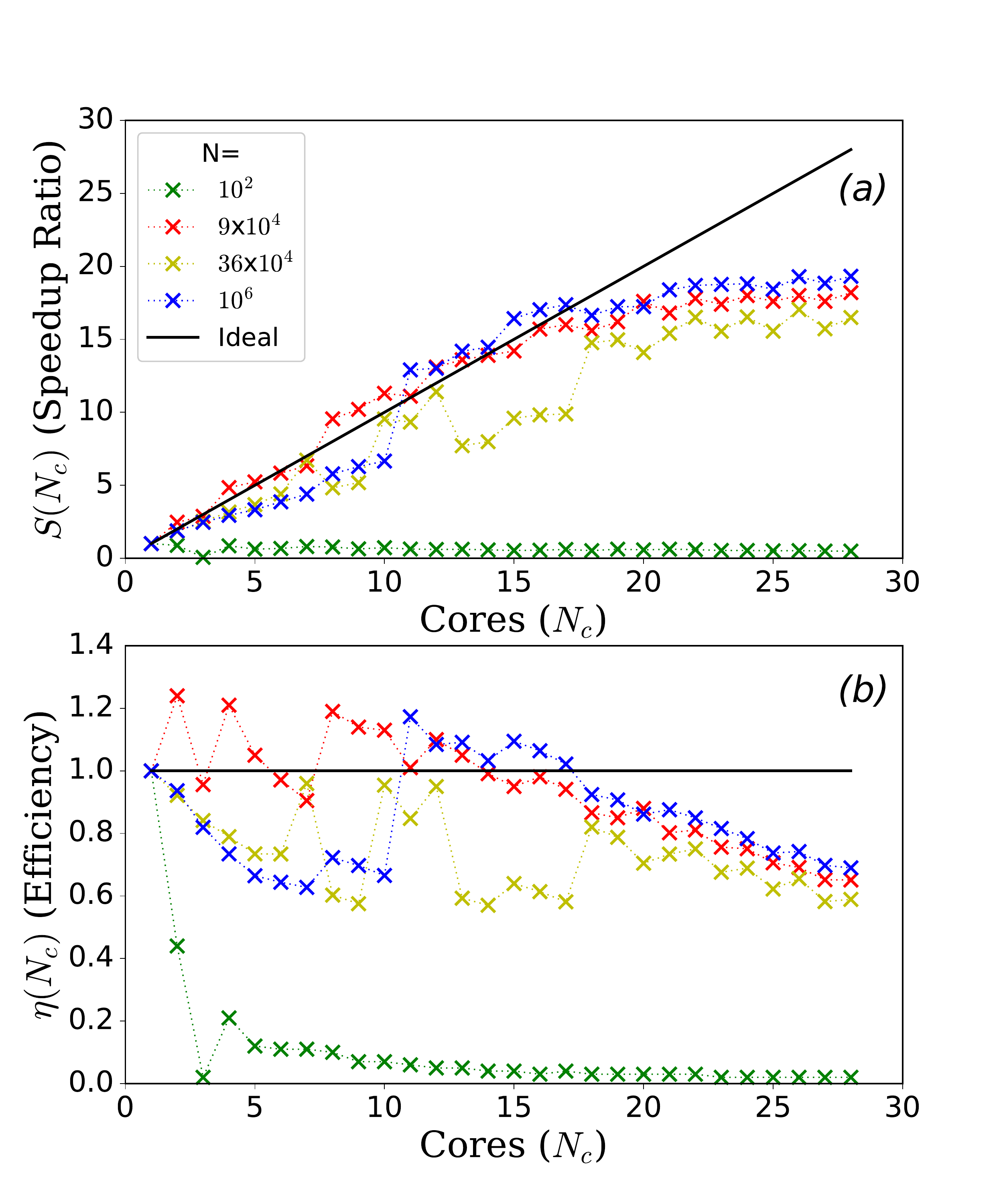}
    \caption{Computed (a) Speedup ratio and (b) Efficiency of Krylov-Schur method for varying transient states ($N = \mathrm{L}\times \mathrm{L}$) as a function of cores ($N_c$).}
   \label{fig:sp_eff}
\end{figure}

In addition to this, we also calculated the speedup ratio and efficiency, defined respectively by  
\begin{equation}
    S(N_c) = \frac{T(1)}{T(N_c)}, \qquad \eta = \frac{S(N_c)}{N_c}, 
\end{equation}
where $N_c$ corresponds to the number of cores. Again, the tolerance parameter was set to $10^{-10}$ to provide a reasonable trade-off between accuracy and performance. We display in Fig.~\ref{fig:sp_eff} the speedup ratio and efficiency for the following set  of transient states numbers: $N = 10^2, 9\times 10^4, 36 \times 10^4$ and $10^6$. The scalability efficiency $\eta(N_c)$ decreases as the number of cores increases. However, as the number of transient states increases, the efficiency increases. As shown in Fig.~\ref{fig:sp_eff}, we observe that there is a deviation from the ideal linear speedup for $N_c$ greater than 15. This results from the loss of communication between the processors. In the following, we found a similar scalability for the more realistic model discussed next and that a sequential implementation of the preconditioned KS solver was sufficient to solve our eigenvalue problems.

\subsection{Survival Probabilities}\label{sp_results}
We further assess the efficiency of the various subspace projection algorithms in computations involving the absorption of a vacancy by a cavity in aluminum. 
As in the previous two-dimensional model system, the tri-dimensional lattice cell is periodically replicated, entailing that an array of cavities is indeed involved. The cavity concentration is determined by the cell lengths of the lattice. This second model is more realistic as it accounts for the deformation field created by the spherical cavity whose radius is $20.7 \AA$. Temperature is $60$~K. 
The evolution operator given by Eq.\eqref{eq:prob_exp_tra} provides us with the survival probability distribution Eq.\eqref{eq:survival_probability} at time $t$. The goal is to characterize the absorption kinetics of a single vacancy around a cavity at 600K. The coordination number ($Z=12$) for the FCC lattice indicates the degree of matrix sparsity, i.e., a maximum of $Z+1$ elements per row are non zero. This model consists of 34801 transient states. We consider that the initial probability distribution of the vacancy is either localized or uniform: in the first setup, the vacancy is located at a distance of 57.98 \AA ~from the cavity center along the $\big[ 110 \big]$ direction. The second setup reflects the homogeneous creation of vacancies under irradiation, neglecting spatial correlations originating from cascades. In KMC simulations, the defect position would be drawn from the uniform multinomial law. 

The survival probability distribution are computed using both setups of initial distributions, resorting to Eq.\eqref{eq:survival_probability}. Again, the tolerance parameter was set to $10^{-12}$ and computations are performed using SLEPC and PETSC softwares \cite{slepc-users-manual,slepc-toms,str-6,petsc-efficient}.

\begin{figure}[!ht]
    \centering
    \includegraphics[width=0.5\textwidth]{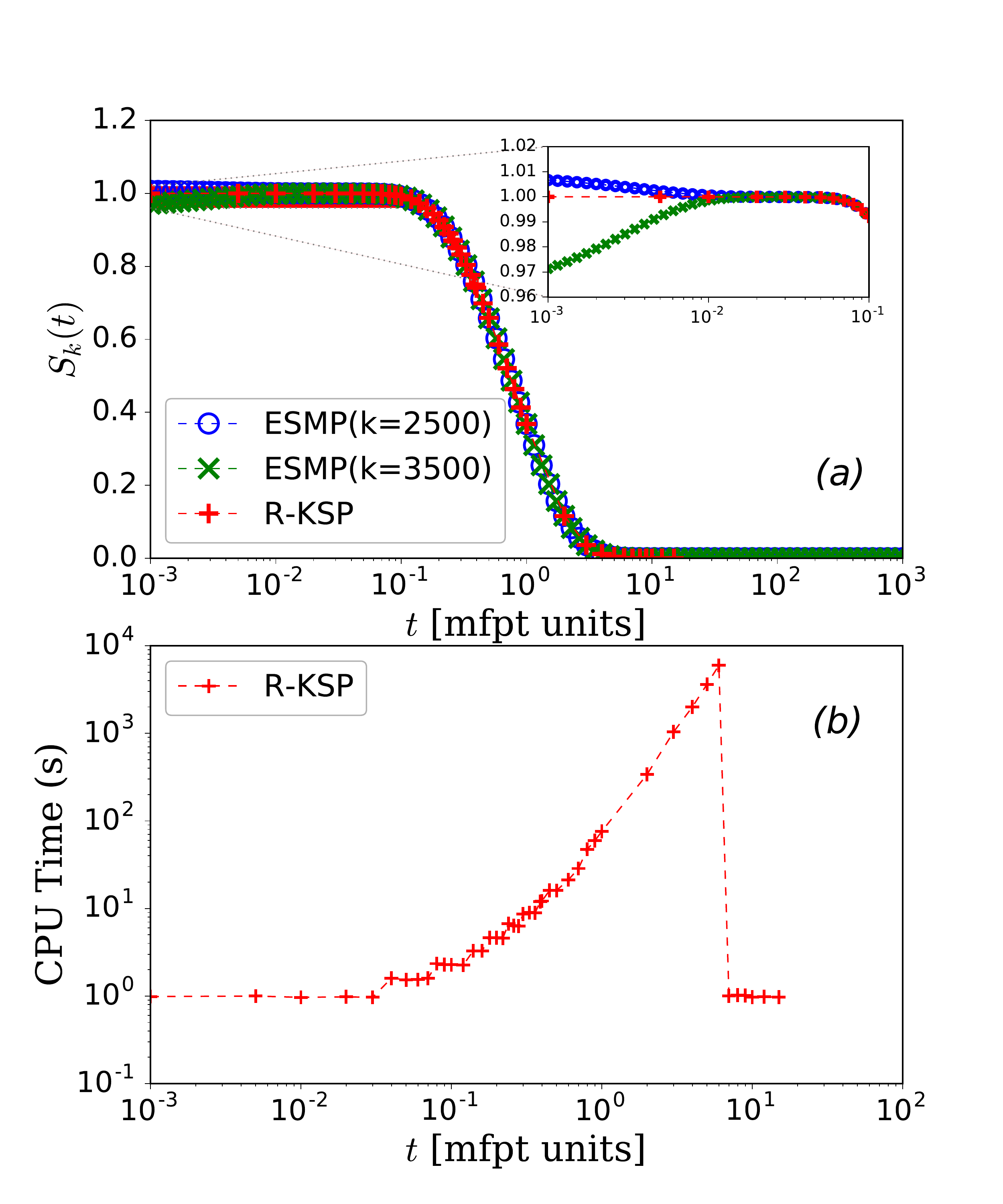}
    \caption{(a) Survival probability distribution computed using ESMP (2500 and 3500 eigenvalues), and R-KSP plotted in blue, green, and red, respectively. (b) CPU cost for computing survival probability vector at a particular instant of time using R-KSP.}
   \label{fig:rksp_profiles}
\end{figure}

We display in Fig.~\ref{fig:rksp_profiles}(a) the survival probability distribution computed using the R-KSP method~[\ref{rksp}] as well as the ESMP method [Section~\ref{esmp}] for up to the 3500 eigenvalues. For the ESMP method, we extracted the lowest eigenvalues by implementing the Krylov-Schur method with the tolerance parameter set to  $10^{-12}$. For the R-KSP method, the tolerance parameter set to  $10^{-10}$ for better convergence. We extracted these lowest eigenvalues by implementing the Krylov-Schur method. The displayed distributions are scaled using the MFPT, estimated independently using the two standard linear solvers (sparse Cholesky and conjugate gradient). As it is observed from Fig.~\ref{fig:rksp_profiles}(a), the initial survival probability obtained from the ESMP method is not equal to one. This discrepancy is attributed to the fact that a substantial number of eigenmodes, higher than 3500, significantly contributes to the short time kinetics. The cost of extracting a huge portion of the eigen spectrum limits the applicability of the ESMP method.  As for the R-KSP method, the survival probability is equal to one, but one needs to specify the Krylov subspace dimension. Typically, the value for the dimension of the Krylov subspace should not be less than 100. The number of restarted iterations to converge depends on the subspace dimension. 

We display in Fig.~\ref{fig:rksp_profiles}(b) CPU times versus physical times for the R-KSP method. We observe that the simulation requires only 0.9 seconds for short time kinetics. However, this method becomes less efficient as time increases. It requires four more orders of magnitude of CPU time to converge at times larger than MFPT. It entails that R-KSP can be practically implemented for the evaluation of survival probabilities at short times only. We now investigate the range of applicability of KSMP and EKSMP methods and check whether they exhibit the same limitations as the ESMP method. The R-KSP method provides the reference data to support the EKSMP and KSMP algorithms.

\begin{figure}[hbt!]
     \centering
     \includegraphics[width=1.0\textwidth]{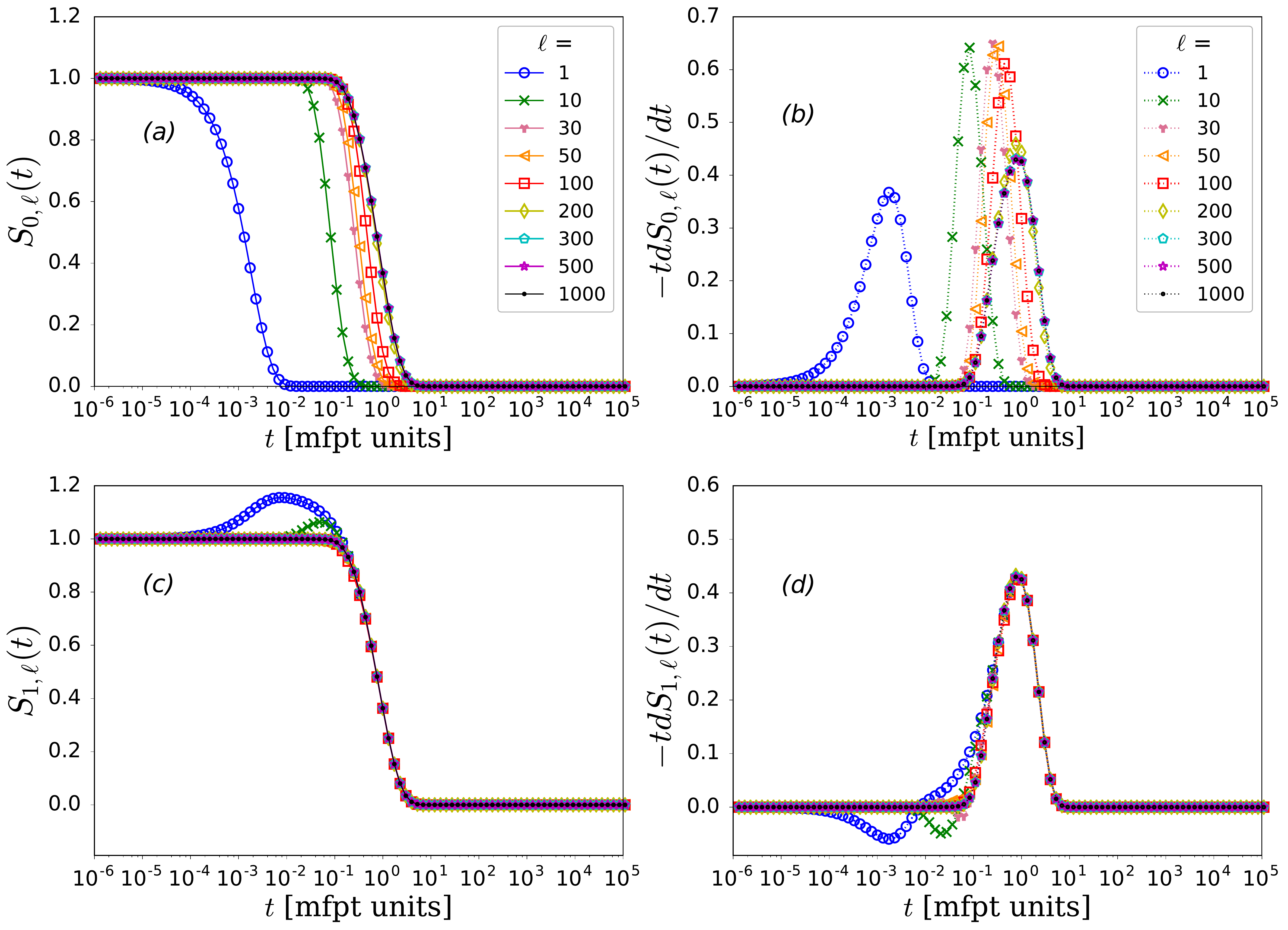}
     \caption{Comparison of survival probabilities and first passage distributions evaluated for the localized initial distribution using the KSMP method is presented in panels (a,b), and the EKSMP method is presented in panels (c,d).}
     \label{fig:non_uniform_distribution}
\end{figure}

\begin{figure}[!ht]
     \centering
     \includegraphics[width=1.0\textwidth]{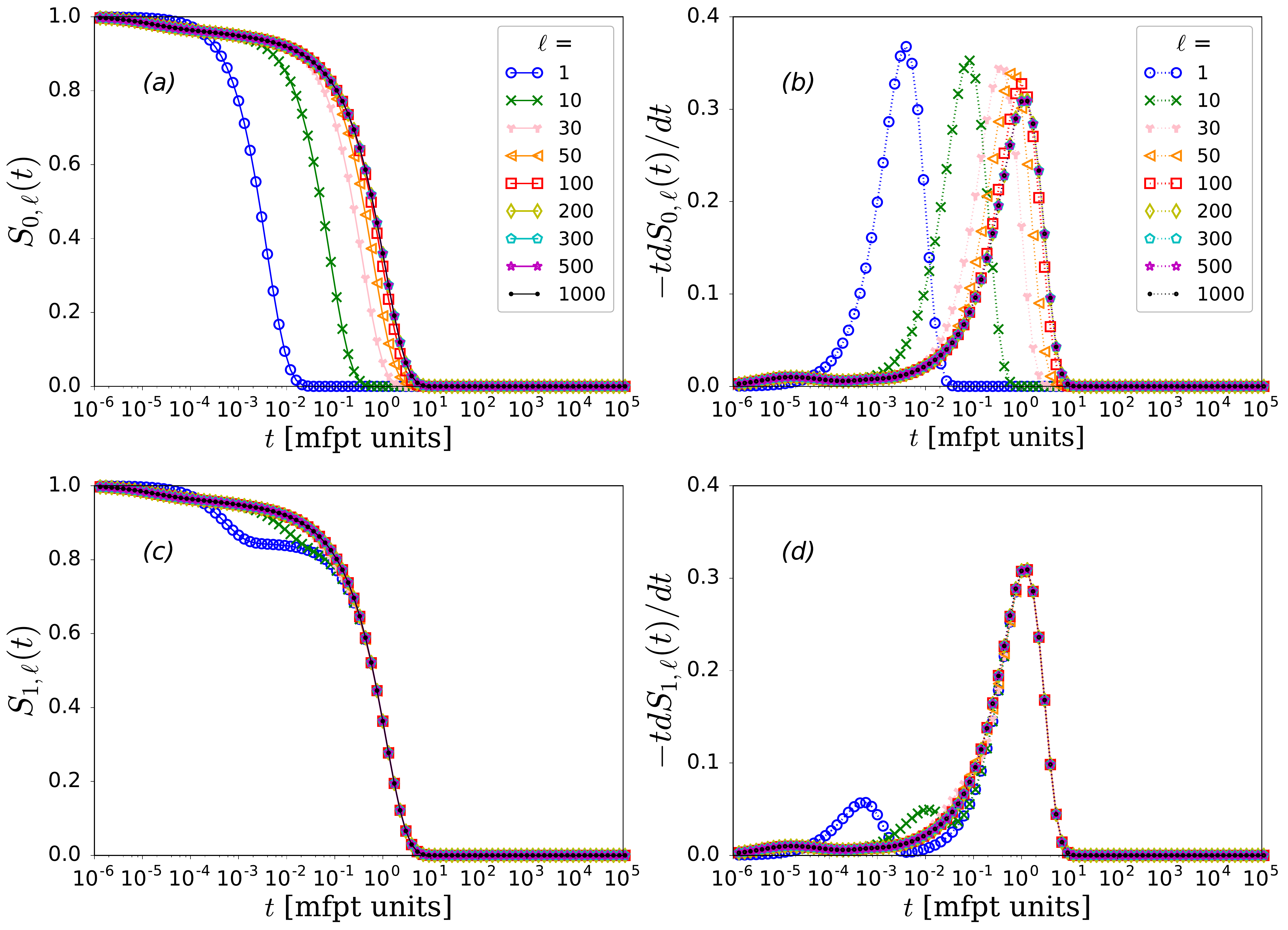}
     \caption{Comparison of survival probabilities and first passage distributions evaluated for the uniform initial distribution using the KSMP method is presented in panels (a,b), and the EKSMP method is presented in panels (c,d).}
     \label{fig:uniform_distribution}
\end{figure}

The survival probability distributions estimated using KSMP and EKSMP methods for localized initial distribution are displayed in Fig.~\ref{fig:non_uniform_distribution}. The full eigenvalue decomposition of the reduced matrix $\mathbf{T}_\ell$ was performed using a dense solver from LAPACK library. The Krylov subspace dimension $\ell$ is varying. We observe a fast convergence with respect to the Krylov subspace dimension for EKSMP and KSMP methods. Interestingly, the EKSMP method with $k=1$ requires a substantially smaller Krylov subspace, about 50 to 100, than the KSMP method does, about 500. This implies that the extra dimension required by the KSMP method aims at capturing the long-term kinetics of the QSD mode. This trend is more pronounced when the initial distribution is uniform, as observed in Fig.~\ref{fig:uniform_distribution} wherein the survival probability and first passage distribution are displayed using KSMP and EKSMP methods and the same setups. This feature is attributed to the higher overlap between the initial distribution and the QSD. The dependence on initial conditions is the main limitation of EKSMP and KSMP methods, compared with the ESMP method that is less sensitive to the initial probability distribution. The significant number of non-local eigenmodes that must be computed contribute to the short time kinetics for a whole range of initial conditions. ESMP is, however, much more costly in terms of memory and CPU, as shown hereafter. 


The CPU time taken by the three methods to calculate the survival probability distributions for the absorption problem are reported in Table~\ref{tab:times}. We used two different processors, referred to as processors A an B and whose features are given in the Table's caption. Sequential runs were performed with processors A and B. Parallel runs of ESMP method were also performed with processor A with 28 cores, yielding a speedup by a factor of 15 similar the one reported in Fig.~\ref{fig:sp_eff} for the parallel simulations of the two-dimensional model.  
\begin{table}
    \centering
    \begin{tabular}{|c|c|c|c|c|}
          \hline
           \textbf{Processor} & \textbf{Methods} & \textbf{Condition} & \textbf{CPU Time (s)} & \textbf{Cores ($N_c$})  \\[0.8ex]
           \hline\hline
           \multirow{2}{4em}{Proc-A}  &\multirow{2}{*}{ESMP}  & $k=3500, \ell$=0    & 3.02 $\cdot 10^4$ & 1  \\
                                      &                       & $k=3500, \ell$=0    & 2.13 $\cdot 10^3$ & 28 \\
           \hline
           \multirow{9}{4em}{Proc-B}  &\multirow{2}{*}{ESMP}  & $k=1,   \ell$=0     & 8.93 $\cdot 10^1$  & 1 \\
                                      &                       & $k=3500,\ell$=0     & 2.96 $\cdot 10^4$ & 1 \\  
           \cline{2-5}
                                      &\multirow{2}{*}{KSMP}  & {$k=0, \ell$=50}    & 1.10 $\cdot 10^1$& 1  \\ 
                                      &                       & {$k=0, \ell$=500}   & 8.60 $\cdot 10^1$& 1  \\
           \cline{2-5}
                                      &\multirow{2}{*}{EKSMP} & {$k=1, \ell$=50}    & 9.83 $\cdot 10^1$& 1  \\ 
                                      &                       & {$k=1, \ell$=500}   & 2.35 $\cdot 10^2$& 1  \\
           \cline{2-5}
                                      &ESMP-CP  & {$k=3500,\ell$=0} & 4.66 $\cdot 10^3$ & 1  \\ 
                                      &EKSMP-CP & {$k=1,\ell$=500}  & 1.70 $\cdot 10^2$ & 1   \\
          \hline                
     \end{tabular}
    \caption{CPU time (CPU) taken by ESMP, KSMP, and EKSMP methods to compute the first passage distributions for the single vacancy absorption. Processor A is Intel(R) Xeon(R) Gold 6132 CPU running, each node running at 2.60GHz with $2\times14$ cores. Processor B is Intel i5-8400H running at 2.5GhZ with eight cores. CP stands for Cholesky preconditioning using MUMPS package and $N_c$ is the used number of cores. }
    \label{tab:times}
\end{table}
The ESMP method is computationally more expensive than EKSMP and KSMP methods. Furthermore, it is observed that EKSMP takes more CPU time than KSMP method does. It is because EKSMP method must extract the slowest eigenmode corresponding to the QSD, which requires 153 seconds of computational time. However, if one is interested in the absorption kinetics from a set of initial vacancy positions, then the EKSMP method will be more efficient because the QSD is computed once.

\begin{figure}[!ht]    
    \centering
    \includegraphics[width=0.8\textwidth]{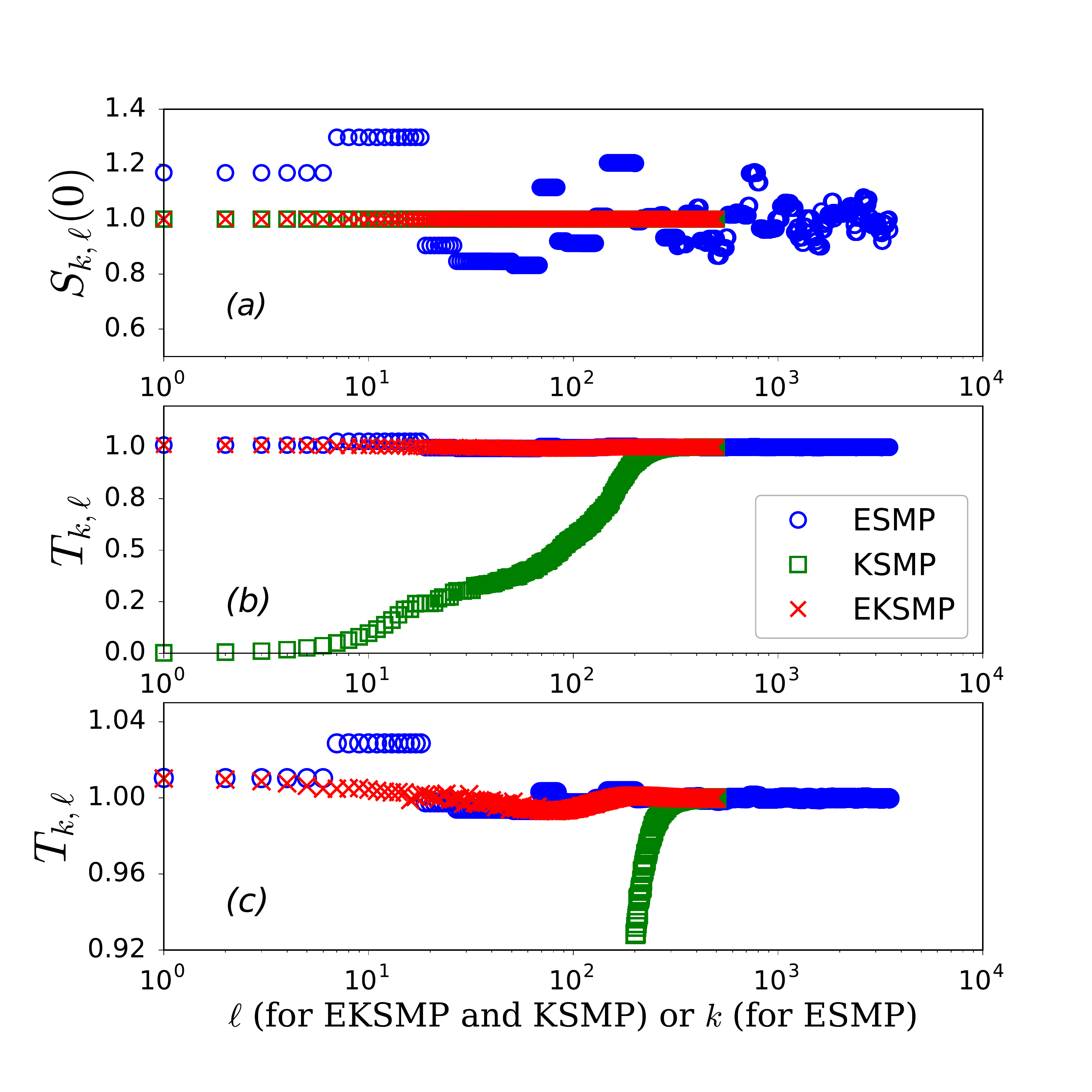} 
    \caption{Comparison of (a) survival probabilities $S_{k}(0)$, (b) reduced MFPT $T_{k,\ell}$ computed for ESMP (3500 eigenvalues), KSMP (500 eigenvalues) and EKSMP (500 eigenvalues) methods. Panel (c) zooms over values of the reduced MFPT ranging from 0.92 to 1.04 for the respective methods.}
    \label{fig:sk_tk_methods_compa}
\end{figure}

We quantified the reduced MFPT $T_{k,\ell}$ and survival probability $S_{k,\ell}(0)$ estimated using the different methods to monitor the convergence as a function of $k$ and $\ell$. The results are displayed in Fig.~\ref{fig:sk_tk_methods_compa}. The blue curve corresponds to the ESMP method. The 3500 eigenvalues evaluated previously have been used. The green and red curves correspond to EKSMP and KSMP for $\mathit{l} = 500$, respectively. The reduced MFPT converges within $2\cdot 10{-3}$ after $\ell=200$ for EKSMP. Truncation errors for estimating the MFPTs are lower with a monotonous behavior and a fast convergence for the KSMP and EKSMP methods as compared to the ESMP method. In panel (c) of Fig.~\ref{fig:sk_tk_methods_compa}, we clearly distinguish the convergence rates of the three methods. 

\begin{figure}[!ht]
    \centering
    \includegraphics[width=0.65\textwidth]{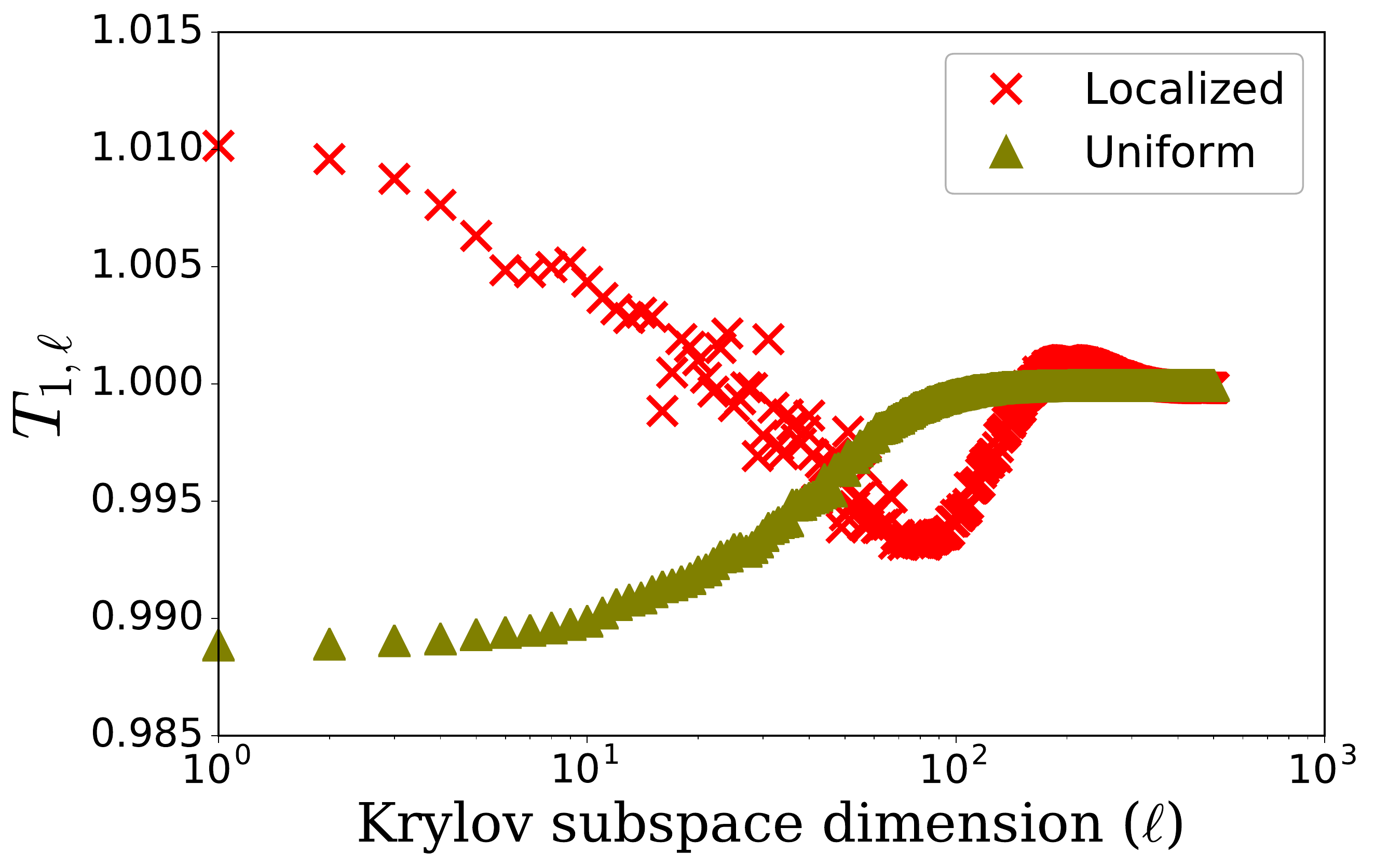} 
    \caption{Reduced and truncated MFPT $T_{1,\ell}$ evaluated for a localized and uniform initial distribution using EKSMP method.}
    \label{fig:sk_tk_uni_nonuni}
\end{figure}

To sample the MFPT starting from a uniform initial distribution, we used the EKSMP method. The quantity $T_{k,\ell}$ was computed with $k=1$, and the results are displayed in Fig.~\ref{fig:sk_tk_uni_nonuni}. Convergence is clearly much faster when the initial distribution is uniform than when it is localized. Hence, a smaller Krylov subspace is required to evaluate the first passage distributions with accuracy and at a low computational cost. The main argument for using both Krylov subspace and eigenvector subspace projections is to reduce the dimension of the former subspace and to reuse the second subspace in other calculations. In~\ref{appendix:setup}, additional results are reported concerning the use of Cholesky preconditioning (CP) and about the relevance of increasing the eigenvalue subspace $k$ in EKSMP method. It is shown that CP decreases the overall CPU cost and facilitates the extraction of additional eigenpairs. It should therefore be enabled whenever possible. 

\subsection{Vacancy absorption kinetics}\label{vak}
To visualize the absorption kinetics, we compute and display the probability fluxes to the cavity and the sink strengths from the initial sites of the mobile vacancy. The goal is to investigate the effect of the elastic deformation on the vacancy pathway to cavity. 
\subsubsection{Vacancy flux to cavity}\label{rtv_results}
\begin{figure}[!h]
    \centering
    \includegraphics[width=1.0\textwidth]{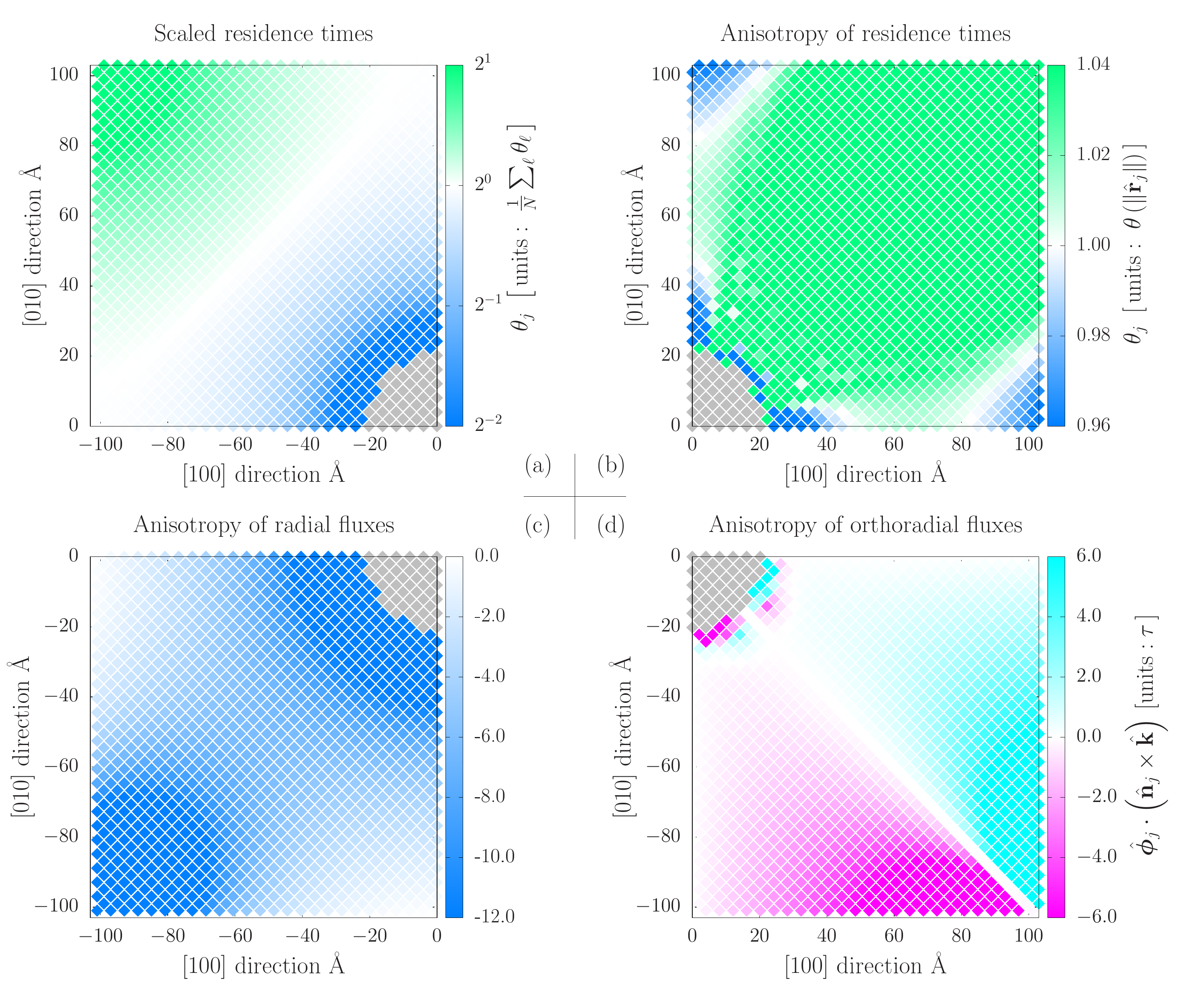} 
    \caption{Estimation of residence times (a,b), radial vacancy flux (c), and ortho-radial vacancy fluxes (d) starting from a localized initial distribution. Absorption of a single vacancy initially located at a distance of 57.98~\AA~from the cavity center in $<100>$ crystalline directions. Coordinates of displayed sites satisfy $\hat{\mathbf{r}}_j\cdot \hat{\mathbf{k}}=0$ where $\hat{\mathbf{k}}$ is the normalized basis vector orthogonal to $(001)$.}
    \label{fig:figx}
\end{figure}

We first computed the mean residence time vector $\bm{\theta}$ defined by 
\begin{equation}\label{res_vec}
    \bm{\theta}^T=\bm{\pi}^T \left(\absorb^\mathrm{a}\right)^{-1}, 
\end{equation}
for initial distribution $\bm{\pi}$ by casting this equation in the form of Eq.~\eqref{eq:ksmp_ax} and using both the sparse Cholesky and CG solvers (to check that results were matching). Then, introducing the three dimensional lattice coordinates $\hat{\mathbf{r}}_j$ of the vacancy for state $j$~\cite{athenes_elastodiffusion_2019}, the vacancy flux was computed from the relation
\begin{equation}\label{current_density}
    \hat{\bm\phi}_{j} = \frac{1}{2 v} \sum_{\ell} \big( \theta_j K_{j\ell} - \theta_{\ell} K_{\ell j} \big) \left(\hat{\mathbf{r}}_{\ell}-\hat{\mathbf{r}}_{j}\right)
\end{equation}
where $\theta_j K_{j\ell} - \theta_{\ell} K_{\ell j}$ are the mean probability currents between both the transient and absorbing states, $\bm{\theta}$ is the mean residence time vector and $v$ represents the unit cell volume assumed to be uniform over the simulation box. All panels in Fig.~\ref{fig:figx} represents a quarter of the $(0 0 1)$ plane containing the center of the cavity. The vacancy resides on the $(100,100,0)$ Cartesian coordinates in Fig.~\ref{fig:figx}, along $<110>$ direction. We computed residence times Eq.\eqref{res_vec} and vacancy fluxes Eq.\eqref{current_density} for the localized initial distribution for sites $j$ using linear solver. The algorithm used to compute these quantities are detailed in~\cite{athenes_elastodiffusion_2019}. Figure~\ref{fig:figx}.(a) represents the scaled residence times. We observe that the residence time is high at the periphery and low near the center, where the vacancy is more easily absorbed. Besides, anisotropy in the residence times can be observed as the vacancy evolves through the system and stays for shorter times along $[100]$ and $[010]$ directions, as observed in Fig.~\ref{fig:figx}(b). The anisotropy in radial fluxes can also be observed in Fig.~\ref{fig:figx}(c). Absorption path along the crystalline direction $[110]$ depicts the anisotropic behavior.

\begin{figure}[!h]
    \centering
    \includegraphics[width=1.0\textwidth]{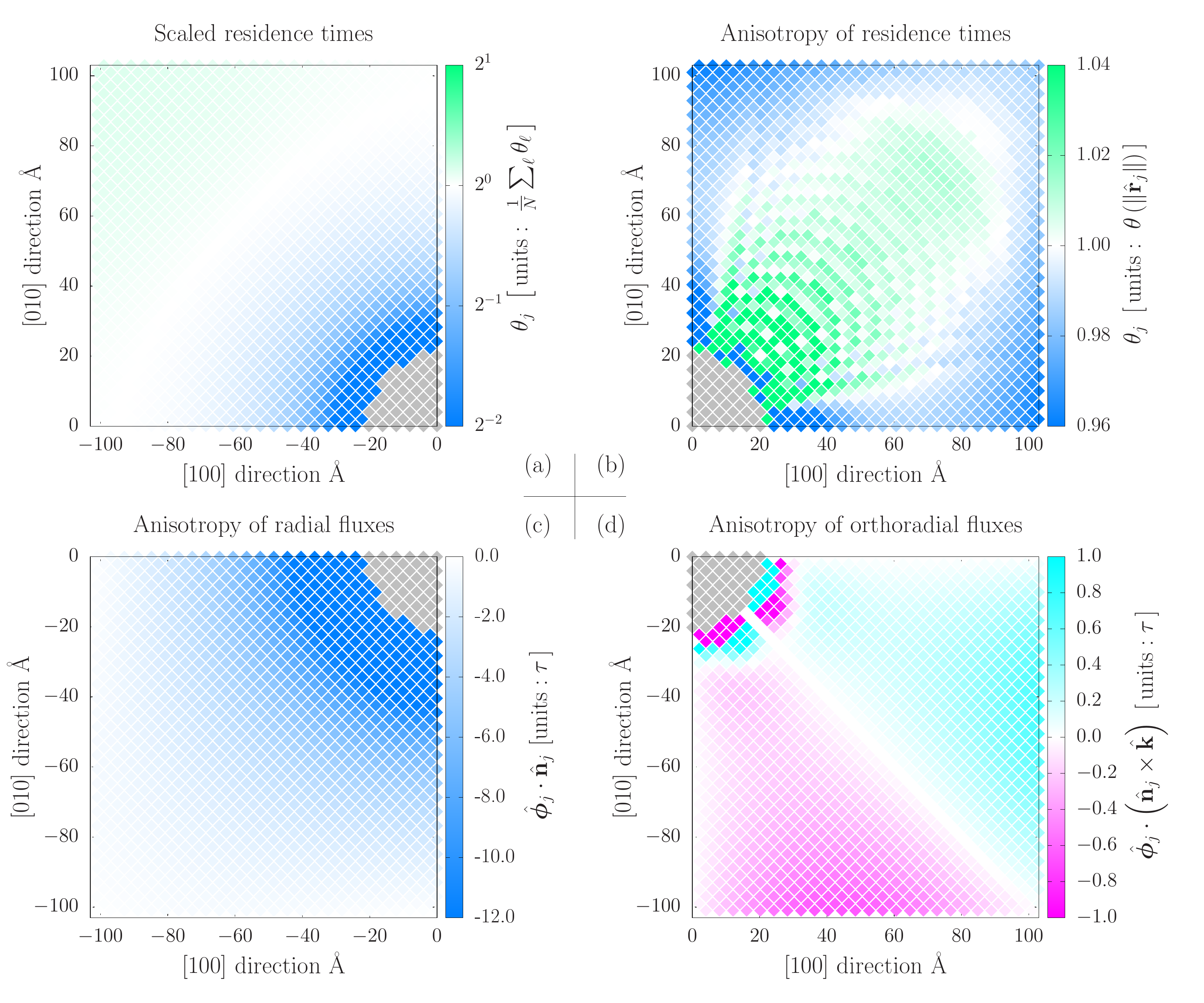}
    \caption{Estimation of residence times (a,b), radial vacancy flux (c), and ortho-radial vacancy fluxes (d) using a uniform initial distribution. The cavity center is in $<100>$ crystalline directions. Coordinates of displayed sites satisfy $\hat{\mathbf{r}}_j\cdot \hat{\mathbf{k}}=0$ where $\hat{\mathbf{k}}$ is the normalized basis vector orthogonal to $(001)$.}
    \label{fig:figu}
\end{figure}

The computed vacancy fluxes and residence times for the uniform distribution are displayed in Fig.~\ref{fig:figu}. The trends are qualitatively similar to those observed when the initial distribution was localized, but not quantitatively. In this setup, the residence times for vacancy at each site are less, implying that vacancy absorption happens faster. 

\begin{figure}[!ht]
    \centering
    \includegraphics[width=1.0\textwidth]{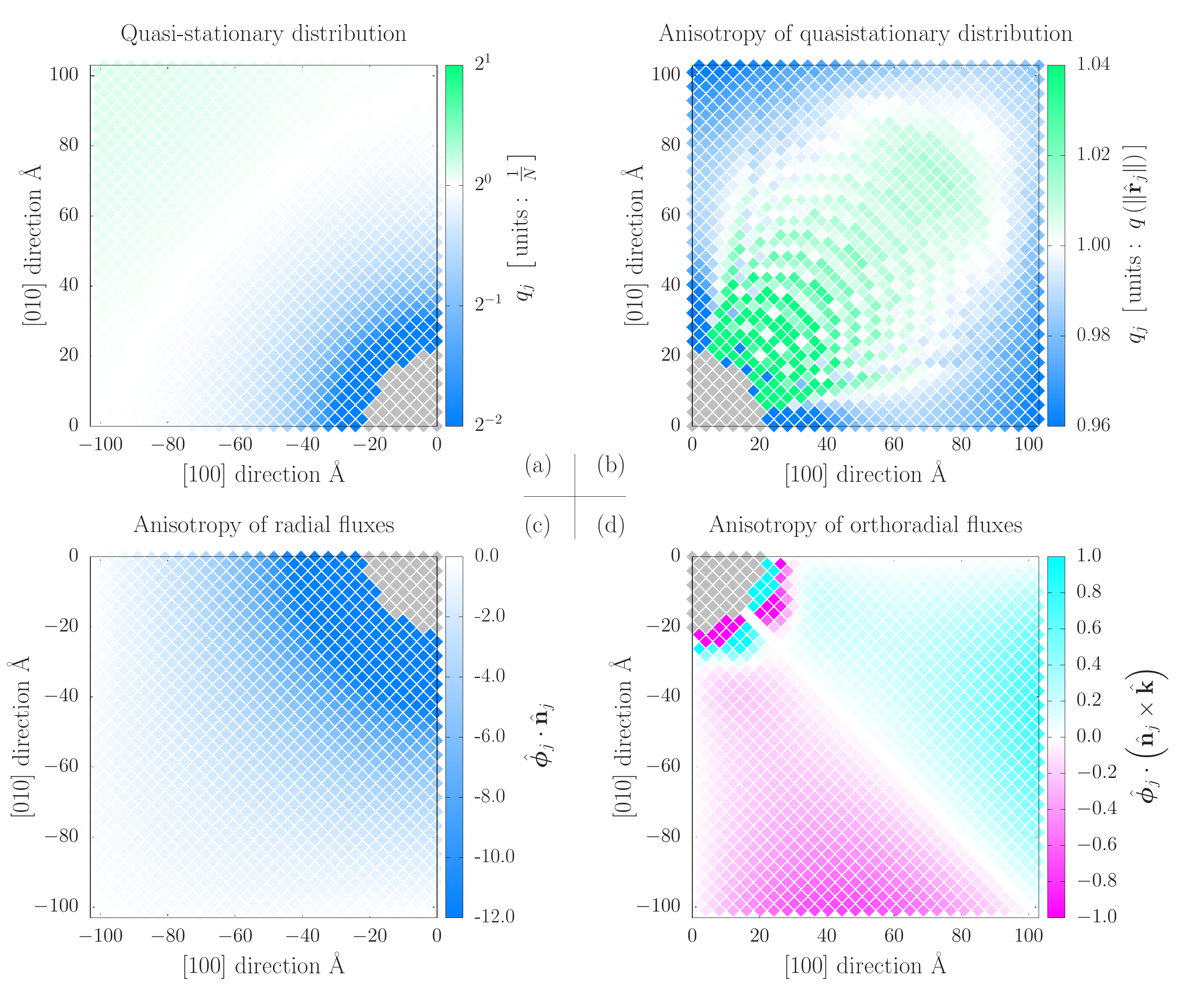} 
    \caption{Estimated quasi-stationary distribution (a), anisotropy of quasi-stationary distribution probability (b), anisotropy of radial vacancy fluxes (c), and anisotropic ortho-radial vacancy fluxes (d) for the localized initial distribution.}
    \label{fig:figqsd}
\end{figure}

Replacing the reduced residence time vector $\bm{\theta}/\bm{\theta}^T\ones$ with the  quasi-stationary probability vector $\mathbf{q}$ in Eq.\eqref{current_density}, provides the fluxes in the asymptotic time limit. The results are displayed in Fig.~\ref{fig:figqsd}. The radial dependence of the QSD is shown in Fig.~\ref{fig:figqsd}(a). The reduced probability decreases from 2 far from the cavity to 0.25 at the cavity periphery where the vacancy is about to get absorbed. The anisotropic nature of the radial and ortho-radial fluxes can also be observed in Fig.~\ref{fig:figqsd} (c,d). The more pronounced anisotropic behavior observed in Fig.~\ref{fig:figx} is due to the localized initial distribution, whereas anisotropy is less critical in Fig.~\ref{fig:figu} and Fig.~\ref{fig:figqsd}. The redidual anisotropy associated with the QSD is entrirely due to the presence of the elastic field created by the cavity. 

\subsubsection{Sink strengths}\label{ssf_results}
\begin{figure}[!ht]
    \centering
    \includegraphics[width=0.7\textwidth]{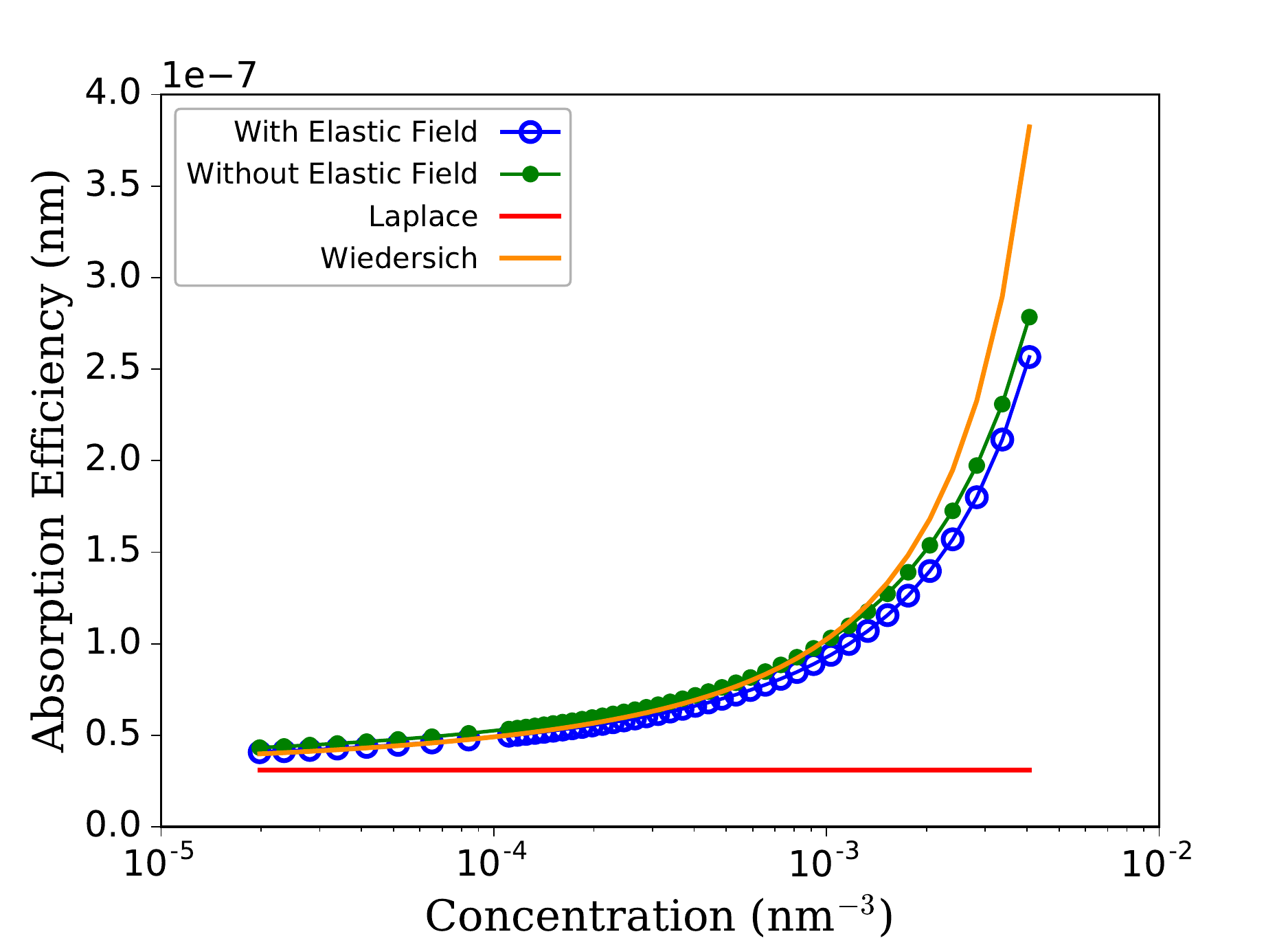} 
    \caption{Comparison of absorption efficiencies calculated for the sink absorption model using MFPT from Eq.\eqref{def:mfpt_1} and uniform initial distribution to those obtained from Laplace equation~\eqref{eq:laplace} and Wiedersich equation~\eqref{eq:wiedersich}. }
    \label{fig:absorption_coeff_uni}
\end{figure}

We next quantify the sink strengths defined by
\begin{equation} \label{eq:sink_strength}
     k^2 = \frac{1}{\tau D_v} = \frac{N}{\sum_{j=1}^N\tau_j^{(N)} D_v} 
\end{equation}
where $\tau$ denotes the MFPT associated with the uniform distribution and $D_v$ the diffusion coefficient of the vacancy. Letting $\nu$ denote the vacancy-atom exchange frequency, we have $D_v = a^2 \nu$ and $\nu = \nu_0 e^{-E^\mathrm{m}/(k_\mathrm{B}T)}$ with $E^{\mathrm{m}}$ the migration energy, $\nu_0$ the Debye frequency of aluminum, $k_\mathrm{B}$ Boltzmann's constant, and $T$ the temperature (600 K). To compare between simulations and theory, it is convenient to define absorption efficiencies by renormalizing the sink strengths by the cavity concentration $C_c$: 
\begin{equation} \label{eq:absorption}
    \kappa =  k^2/C_c. 
\end{equation}
We used Laplace and Wiedersich theoretical models~\cite{wiedersich1972theory} to compare the absorption efficiency estimated using our simulation model. The Laplace model is described as
\begin{equation} \label{eq:laplace}
    \kappa = 4\pi r_{cv}
    \end{equation}
where $r_{cv}=r_{ca}+r_{va}$ is the sum of the cavity radius and the vacancy radius. The Wiedersich model depends on the sink concentration and is given by~\cite{wiedersich1972theory}
\begin{equation} \label{eq:wiedersich}
    \kappa = 4\pi r_{cv} \frac{1-\eta^3}{1-\frac{9}{5}\eta + \eta^3 - \frac{1}{5}\eta^6},
\end{equation}
where $\eta$ is $r_{cv}/R$ and $R$ is the average half distance between sinks (cavity). The distance $R$ is calculated by $R=\sqrt[3]{C_c}/2$~\cite{carpentier_effect_2020}. The computed absorption efficiencies are displayed in Fig.~\ref{fig:absorption_coeff_uni}. The curve provided by Wiedersich model Eq.\eqref{eq:wiedersich} have an almost perfect match with the simulation results. The constant value given by Laplace model Eq.\eqref{eq:laplace} corresponds to absorbing efficiency in the limit of zero concentration of sinks. The absorption efficiency obtained in the simulations seem to converge to this value as concentration decreases. Another interesting feature obtained by the simulation is the negligible dependence of sink strengths on elastic interactions even for high concentration of the cavity, at least for the uniform initial distribution used in this work. 

\begin{figure}[!ht]
    \centering
    \includegraphics[width=0.7\textwidth]{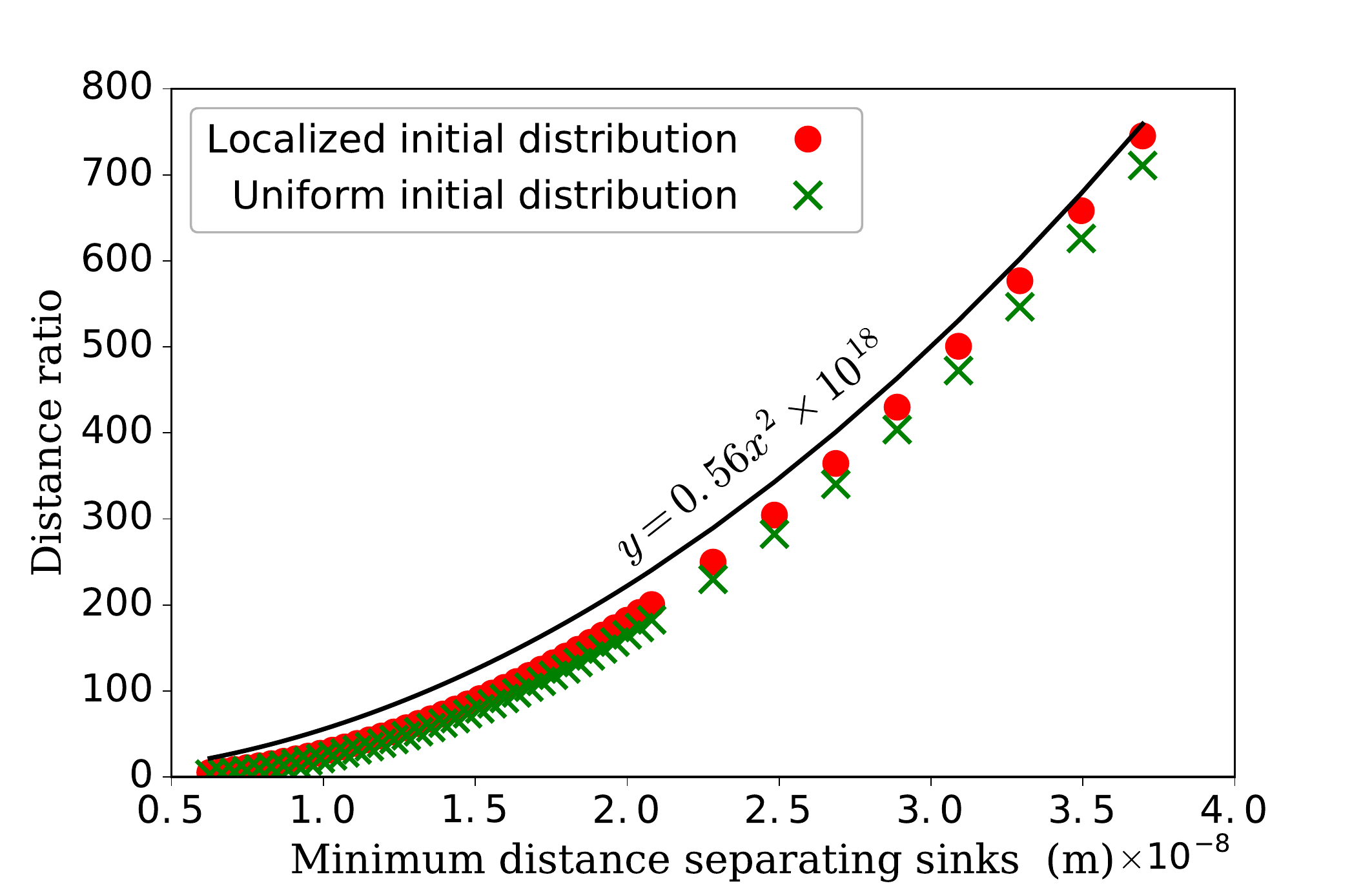} 
    \caption{Distances traveled by the vacancy before absorption scaled by the minimum distance separating the sinks, for uniform and localized initial distributions, and plotted as a function of the separating distance. }
    \label{fig:sink_distance}
\end{figure}

We note that the proposed algebraic approach to sink strength computations possibly gives access to a lot of additional information, such as the total number of vacancy jumps and the distance traveled by the vacancy before absorption by the cavity, and also the dependence of the sink strengths on the initial probability distribution. The transient number of vacancy jumps is deduced from the residence time vector $\bm{\vartheta}^T=\bm{\pi}^T\left(\mathbf{A}^\mathrm{a}\right)^{-1} $, a quantity depending on the initial probability vector $\bm{\pi}$. The number of transient jumps from any state being equal to the product of the residence time and the jump frequency, the total number of jumps is obtained by summing over all transit and writes $\sum_i \vartheta_i A^\mathrm{a}_{ii}$. Multiplying by $\mathrm{a}\sqrt{2}/2$, the nearest neigbour distance between lattice site yeilds the distance traveled by the vacancy prior absorption. The later distance scaled by minimum distance separating the cavities, i.e. the period length, has been displayed in Fig.~\ref{fig:sink_distance} for a range of cavity concentrations.  We observe that the traveled distance increases much faster than the separating distance. The distance ratio follows as a parabolic law, entailing that the traveled distance grows as the cube of the separating distance. 

\begin{figure}[!ht]
    \centering
    \includegraphics[width=0.8\textwidth]{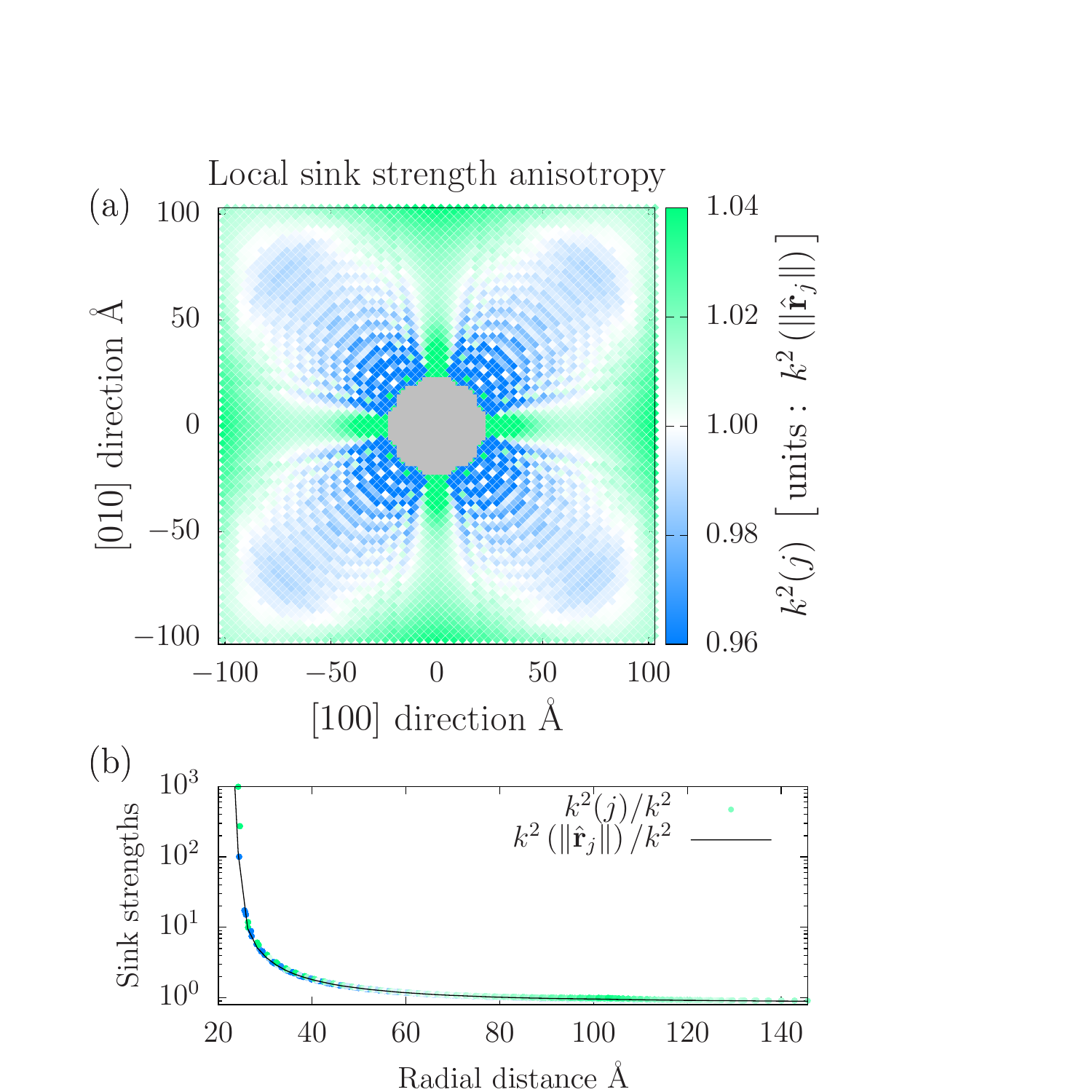}
    \caption{Using uniform distribution, (a) estimation of anisotropic sink strengths for the site $j$, and (b) evolution of the sink strengths parameter over a radial distance between the cavity and the vacancy.}
    \label{fig:figsinkanisotropy}
\end{figure}

We eventually quantify the anisotropy of the sink strengths associated with the localized initial distributions $\mathbf{e}_j$. Local sink strenths are defined by $k^2(j)= 1/(\tau_jD_v)$ where $\tau_j$ is the MFPT from site $j$. Their anisotropy, clearly visible in Fig.~\ref{fig:figsinkanisotropy}.a, is very moderate, in contrast to the radial dependence shown in Fig.~\ref{fig:figsinkanisotropy}.b. 

This means that there is no need to account for elasticity to compute sink strengths of small cavities with respect to vacancies in aluminium under irradiation. Note that the size of the simulation box is restricted to $3 \times 10^6$ sites due to memory constraints, and that the cavity sizes is also modest. We however expect a higher effect of the elastic field created by interstitial loops on the absorbing/emitting fluxes of point defects in aluminum~\cite{carpentier_effect_2017}. The anisotropy of these fluxes may introduce a substantial angular dispersion of sink strengths which should ideally be taken into account in cluster dynamics simulations. This can be achieved by implementing the approach developed in a recent work~\cite{carpentier_effect_2020} in which the dispersion effect of the distances between the sinks is correctly accounted for in hybrid cluster dynamics simulations~\cite{terrier_cluster_2017}.

\section{Conclusions} \label{conclusions}
In this work, we have developed several algorithms for characterizing a mobile defect's absorption kinetics by a periodic array of sinks, resorting to the theory of absorbing Markov chains. The sink concentration is determined by the cell size and the absorbing transition rate matrix is deduced from the transition rates between transient states. The initial probability vector is defined by the defect's initial distribution over the transient states. The goal is to predict the time evolution of the state probability vector of the defect until absorption. This entails computing the exponential of the constructed rate matrix, a task ideally accomplished via its full diagonalization. The standard eigensolvers that are used routinely to extract the entire eigenspectrum of low-dimensional matrices fail in high-dimensions due to memory constraints. Fortunately, in many diffusion problems, the involved transition matrices are extremely sparse. The usual approach for sparse high-dimensional matrices is to repeatedly extract portions of the eigenspectrum using iterative solvers based on deflation techniques. However, this still requires a substantial amount of computational time, restricting the range of applicability of this technique for simulating aging kinetics and microstructural evolution in materials science. We have succeeded in overcoming this issue by applying Krylov subspace projection techniques. This approach involves vector-matrix multiplications only and reduces the computational complexity by calculating the exponential of a much lower-dimensional transition rate matrix. Two algorithms have been developed, dubbed KSMP (Krylov Subspace Model Projection) and EKSMP (Eigenvalue and Krylov Subspace Model Projection). KSMP is based on constructing a Krylov basis starting from the initial probability vector and aiming at capturing its subsequent evolution. As the KSMP approach introduces a dependence on the initial vector, we have also developed and tested the additional EKSMP method to deflate the Krylov subspace using the slowest eigenmodes. EKSMP and KSMP methods were implemented to study the absorption kinetics of a vacancy by a cavity in Aluminium. The correctness of the two algorithms was assessed by comparing the results obtained for a subset of times using the Restarted Krylov subspace projection (R-KSP) method as a reference: survival probabilities and first-passage distributions could be accurately reproduced using EKSMP and KSMP methods. Noticeably, an important simplification of the problem to solve stems from the fact that the diffusion process is reversible~\cite{athenes_elastodiffusion_2019} entailing that all transition rate matrices can be symmetrized through diagonal similarity transformations. The reversibility condition is fulfilled in most applications involving the diffusion of defects even though these defects are created by an irreversible process like neutron, ion, or electron irradiation. 

The crucial parameter controlling the convergence of KSMP and EKSMP methods is the dimension of the Krylov subspace. We found that the KSMP method yields accurate results with a Krylov basis whose dimension is five times the cubic root of the matrix dimension (the size of the three-dimensional lattice). Such a dimension for the Krylov basis (KB) makes it possible to capture long sequences of defect hops through the entire cell until absorption, and hence to account for the contribution of the QSD mode. We also observed that for our typical sink problem using the EKSMP method, the KB dimension is considerably reduced, by a factor of 10, even when only the QSD mode is included. This trend results from the fact that the QSD regime is reached very quickly and only involves the local diffusion of the defect. Besides, the extra cost associated with the QSD calculation being less or similar to the cost that is spared by reducing the Krylov subspace dimension, EKSMP method is more advantageous than KSMP method. This is especially true when more than one initial defect distribution is considered, a typical situation occurring in KMC and mean-field applications. Because the eigenvalues are often degenerate and eigenmodes occur in bundles, it was practically inconvenient and computationally expensive to include additional eigenmodes in EKSMP method. The working space dimension must be determined in the Krylov-Schur solver previously selected in our applications for its superior performance compared to other iterative solvers. Overall, we show that it is unnecessary to extract several eigenmodes to characterize the absorbing kinetics fully. 

Concerning sink strength calculations in which the initial defect distribution is homogeneous in space, the convergence with respect to the KB dimension is observed to be faster, as compared to situations with an initially localized vacancy. Overall, we show that Krylov subspace projection methods enable us to study the diffusion of a mobile defect around a sink in model system accounting for elastic dipole interactions and comprising up to million lattice sites. To compute the sink strengths at lower sink densities, the developed Krylov subspace projection technique should be implemented in combination with KMC simulations and conditioning techniques~\cite{athenes:tel-01851686} to circumvent the curse of dimensionality and memory constraints. Finally, we point out that the proposed approach is based on the actual transition rate matrix on the crystal lattice. It may then straightforwardly be applied to investigate the anisotropic migration of interstitial clusters, whose diffusion mechanism mixes fast translations and slow rotations. The approach may allow to compute sink strengths relative any kind of defect clusters exhibiting mixed mobilities efficiently~\cite{adjanor_complete_2020,adjanor_complete_2020-1}.

\section*{Acknowledgements}
Fruitful discussions with Gilles Adjanor, Thomas Jourdan and Jose E. Roman are gratefully acknowledged.

\appendix
\section{Restarted Krylov Subspace Projection}\label{rksp}
\renewcommand{\theequation}{A.\arabic{equation}} 
\setcounter{equation}{0} 
The restarted Krylov subspace algorithm proposed in ref.\cite{eiermann_restarted_2006}, generates Krylov basis of dimension $\ell$. In later step, that algorithm updates the approximation to $f(\absorb)\mathbf{b}$ and discards the basis vectors except the one which serves as an initial vector of the next Krylov subspace \cite{afanasjew_implementation_2008}. The following derivation is to recall the restarted setup using two Lanczos decomposition
\begin{align}
    \absorb\mathbf{V}_\ell^1 = \mathbf{V}_\ell^1 \mathbf{T}_\ell^1 + T_{\ell+1,\ell}^1 \mathbf{v}_{\ell+1}^1 \mathbf{e}_\ell^T \label{eq:rksp_lanczos_decomposition_1} \\
    \absorb\mathbf{V}_\ell^2 = \mathbf{V}_\ell^2 \mathbf{T}_\ell^2 + T_{\ell+1,\ell}^2 \mathbf{v}_{\ell+1}^2 \mathbf{e}_\ell^T \label{eq:rksp_lanczos_decomposition_2}
\end{align}
where $\mathbf{V}^1$ and $\mathbf{V}^2$ are the orthornormal bases of $\mathcal{K}_\ell(\absorb, \mathbf{v}_1)$ and $\mathcal{K}_\ell(\absorb, \mathbf{v}_{\ell+1})$. $\mathbf{T}_\ell^1$ and $\mathbf{T}_\ell^2$ are two tridiagonal matrices. $\mathbf{e}_\ell^T$ denotes the $\ell$th unit coordinate vector $\in \mathbb{R}^{\ell}$. Together the columns of $\mathbf{W}_{2\ell} \coloneqq [\mathbf{V}_\ell^1, \mathbf{V}_\ell^2]$ forms a basis of $\mathcal{K}_{2\ell}(\absorb,\mathbf{b})$. On combining the two Lanczos decomposition Eq.\eqref{eq:rksp_lanczos_decomposition_1} and Eq.\eqref{eq:rksp_lanczos_decomposition_2} to Lanczos-like decomposition we get,
\begin{equation}\label{eq:rksp_comb_lanczos}
    \absorb\mathbf{W}_{2\ell}=\mathbf{W}_{2\ell}\mathbf{T}_{2\ell} + T_{\ell+1,\ell}^2 \mathbf{v}_{\ell+1}^2 \mathbf{e}_{2\ell}^T
\end{equation}
where $\mathbf{T}_{2\ell}$ is the tridiagonal block Matrix represented as,
\begin{equation}\label{eq:tri_diag_mat}
    \mathbf{T}_{2\ell} \coloneqq  \begin{bmatrix}
                                  \mathbf{T}_\ell^1                                      &    \mathcal{O} \\
                                  T_{\ell+1,\ell}^1 \mathbf{e}_1 \mathbf{e}_\ell^T       &    \mathbf{T}_{\ell}^2       
                                  \end{bmatrix}.
\end{equation}
The restarted method of Krylov subspace approximation associated to Eq.\eqref{eq:rksp_comb_lanczos} is given as,
\begin{equation}\label{eq:rksp_f}
    \mathbf{f}_{2\ell} = \beta \mathbf{W}_{2\ell}f(\mathbf{T}_{2\ell})\mathbf{e}_1.
\end{equation}
The $f(\mathbf{T}_{2\ell})$ term exhibits the following block lower triangular structure, 
\begin{equation}\label{eq:rksp_f_tl}
    f(\mathbf{T}_{2\ell}) =  \begin{bmatrix}
                                  f(\mathbf{T}_{\ell}^1)   &    \mathcal{O} \\
                                  \mathbf{X}_{2,1}                  &    f(\mathbf{T}_{\ell}^2)       
                                  \end{bmatrix}.
\end{equation}
Hence, the approximation Eq.\eqref{eq:rksp_f} has the form,
\begin{equation}\label{eq:rksp_final_function}
    \mathbf{f}_{2\ell} = \beta \mathbf{V}_\ell^1 f(\mathbf{T}_{\ell}^1) \mathbf{e}_1 + \beta \mathbf{V}_\ell^2 \mathbf{X}_{2,1} \mathbf{e}_1
\end{equation}
where the first term of Eq.\eqref{eq:rksp_final_function} is evaluated using Arnoldi approximation for the basis $\mathcal{K}_{\ell}(\absorb,\mathbf{b})$. Once the $\mathbf{X}_{2,1} \mathbf{e}_1$ is estimated, the basis vectors of $\mathbf{V}_\ell^1$ are discarded and Eq.\eqref{eq:rksp_final_function} yields the basis of restarting method by updating the Arnoldi approximation. The approximation after $m$ restart cycles is given as
\begin{equation}\label{eq:approx}
    \mathbf{f}^m = \beta \mathbf{W}_{m\ell} f(\mathbf{T}_{m\ell}) \mathbf{e}_1 = \mathbf{f}^{(m-1)} + \beta \mathbf{V}_{\ell}^m [f(\mathbf{T}_{m\ell}) \mathbf{e}_1]_{(m-1)\ell+1:m\ell} 
\end{equation}
where the subscript of the last term in Eq.\eqref{eq:approx} represents the vector with the last $\ell$ components of $f(\mathbf{T}_{m\ell})\mathbf{e}_1$ \cite{eiermann_restarted_2006}. 

\section{Cholesky preconditioning}\label{appendix:setup}
CPU times for performing ESMP or EKSMP simulations are compiled in Tables~\ref{tab:esmp_convergence},~\ref{tab:esmp_mumps_convergence} and~\ref{tab:eksmp_convergence}. From data reported in Table~\ref{tab:esmp_convergence}, we observe that Cholesky preconditioning (CP) should be performed whenever possible because it reduces the overall CPU times and improves the convergence of the KS solver. Eigenvalues being pooled in bundles for symmetry reasons, eigenvectors appears simultaneously in the extraction algorithm. Handling the eigenvalue degeneracy is facilitated by the inverted iterations within CP. 
\begin{table}
    \centering
    \begin{footnotesize}
    \begin{tabular}{c|c|c|c|c|c|c}
        \hline
        \multicolumn{2}{c}{\textbf{Input}} & \multicolumn{5}{|c}{\textbf{Output}} \\ 
        \hline\hline
        \multirow{2}{*}{\textbf{CP}} & \multirow{2}{*}{\textbf{NEV}} & \textbf{CP Time}   & \textbf{EPS Time}  & \textbf{Converged}
        &\multirow{2}{*}{\textbf{NCV}}  & \multirow{2}{*}{\textbf{Result}}\\
        & &  \textbf{(s)} &  \textbf{(s)} & \textbf{Eigenpairs}  
        &                                         &   \\ [0.5ex]
        \hline\hline
        \multirow{12}{*}{\parbox{2cm}{\centering Disabled}}          & 1  & -  & 8.93 $\cdot 10^1$ & 1   & 16 & C\\
                                                                 & 5  & -  & 1.23 $\cdot 10^2$& 1   & 20 & D \\ 
                                                                 & 10 & -  & 1.30 $\cdot 10^2$& 2   & 25 & D \\
                                                                 & 30 & -  & 1.91 $\cdot 10^2$& 21  & 60 & D \\    
                                                                 & 40 & -  & 2.25 $\cdot 10^2$& 31  & 80 & D \\
                                                                 & 50 & -  & 2.35 $\cdot 10^2$& 53  & 100 & C\\
                                                                 & 60 & -  & 1.91 $\cdot 10^2$& 60  & 120 & C\\
                                                                 & 70 & -  & 1.94 $\cdot 10^2$ & 70  & 140 & C\\
                                                                 & 80 & -  & 2.34 $\cdot 10^2$& 82  & 160 & C\\
                                                                 & 90 & -  & 2.45 $\cdot 10^2$& 90  & 180 & C\\
                                                                 & 100 & - & 2.89 $\cdot 10^2$& 102 & 200 & C\\
        \hline\hline                              
        \multirow{12}{*}{\parbox{2cm}{\centering Enabled}}& 1 & 1.58 $\cdot 10^0$ & 1.55 $\cdot 10^1$ &1& 1 & C\\
                                                                             & 5 & 5.53 $\cdot 10^0$ & 1.98 $\cdot 10^1$ & 7& 7& C\\
                                                                      & 10 & 5.92 $\cdot 10^0$ & 1.99 $\cdot 10^1$ & 11  & 11 & C \\
                                                                      & 30 & 1.54 $\cdot 10^1$ & 3.13 $\cdot 10^1$ & 37  & 60 & C \\ 
                                                                      & 40 & 1.94 $\cdot 10^1$ & 3.55 $\cdot 10^1$ & 42  & 80 & C \\
                                                                      & 50 & 2.06 $\cdot 10^1$ & 3.52 $\cdot 10^1$ & 60  & 100 & C\\ 
                                                                      & 60 & 2.20 $\cdot 10^1$ & 3.70 $\cdot 10^1$ & 64  & 120 & C\\
                                                                      & 70 & 2.33 $\cdot 10^1$ & 3.99 $\cdot 10^1$ & 73  & 140 & C\\
                                                                      & 80 & 2.66 $\cdot 10^1$ & 4.37 $\cdot 10^1$ & 81  & 160 & C\\
                                                                      & 90 & 3.05 $\cdot 10^1$ & 4.71 $\cdot 10^1$ & 90  & 180 & C\\
                                                                     & 100 & 3.75 $\cdot 10^1$ & 5.69 $\cdot 10^1$ & 114 & 200 & C\\
                                                                    & 3500 & 1.52 $\cdot 10^3$ & 4.73 $\cdot 10^3$ &3539 & 4000 &C\\
        \hline 
     \end{tabular}
     \end{footnotesize}
     \caption{CPU time taken by ESMP method to extract the indicated number of eigenvalues (NEV) for the vacancy absorption model. The simulations are performed using a single core of an Intel i5-8400H  processor (running at 2.5GhZ). Parameter NCV represents the maximum dimension of the working subspace to be used by the solver. In inputs, Cholesky preconditioning (CP) may be enabled or disabled. In the outputs, C and D of the result column denote whether simulations have converged or diverged, respectively.}
    \label{tab:esmp_convergence}
\end{table}
CP is limited in memory because the computed Cholesky factor is a denser matrix. The largest system CP can solve contains 217245 transient state, as shown in table~\ref{tab:esmp_mumps_convergence}. For the larger systems reported in Table~\ref{tab:eksmp_convergence}, the QSD eigenvector was successively computed using KS solver without preconditioning up to 905681 transient states. 

\begin{table}[]
    \centering
    \begin{footnotesize}
    \begin{tabular}{c|c|c}
         \hline
         \textbf{Cell size (\AA)}  & \textbf{Number of Transient States ($N$)} & \textbf{Time (s)} \\[0.5ex]
         \hline\hline
          66.66 & 14141 & 3.55 $\cdot 10^0$ \\
          74.74 & 21085 & 7.83 $\cdot 10^0$ \\
          86.16 & 34801 & 2.13 $\cdot 10^1$ \\
          98.98 & 53053 & 4.36 $\cdot 10^1$ \\
          107.06 & 68061 & 7.06 $\cdot 10^1$ \\
          127.26 & 116921 & 2.36 $\cdot 10^2$ \\
          139.38 & 154973 & 4.83 $\cdot 10^2$ \\
          147.46 & 184381 & 6.31 $\cdot 10^2$ \\
          151.5 & 200369 & 7.91 $\cdot 10^2$ \\
          159.58 & 217245 & 9.80 $\cdot 10^2$ \\
          \hline
    \end{tabular}
    \end{footnotesize}
     \caption{CPU time taken by ESMP to extract 10 eigenvalues for the indicated number of system size ($N$) using KS solver and Cholesky preconditioning from MUMPS.}
    \label{tab:esmp_mumps_convergence}
\end{table}

\begin{table}[]
    \centering
    \begin{footnotesize}
    \begin{tabular}{c|c|l}
         \hline
         \textbf{Cell size (\AA)}  & \textbf{Number of Transient States ($N$)} & \textbf{Time (s)} \\[0.5ex]
         \hline\hline
          74.74 & 21085 & 4.42 $\cdot 10^1$ \\
          86.86 & 34801 & 1.11 $\cdot 10^2$ \\
          98.98 & 53053 & 2.33 $\cdot 10^2$ \\
          107.06 & 68061 & 4.10 $\cdot 10^2$ \\
          119.18 & 95313 & 5.23 $\cdot 10^2$ \\
          127.26 & 116921 & 7.47 $\cdot 10^2$ \\
          139.38 & 154973 & 1.20 $\cdot 10^3$ \\
          147.46 & 184381 & 1.57 $\cdot 10^3$ \\
          151.50 & 200369 & 1.80 $\cdot 10^3$ \\
          159.58 & 235033 & 2.62 $\cdot 10^3$ \\
          167.66 & 273441 & 3.26 $\cdot 10^3$ \\
          187.86 & 387101 & 5.66 $\cdot 10^3$ \\
          208.06 & 497757 & 9.33 $\cdot 10^3$ \\
          248.46 & 905681 & 2.38 $\cdot 10^4$ \\
         \hline
    \end{tabular}
    \end{footnotesize}
    \caption{CPU time to evaluate survival probability distributions for the indicated number of system size ($N$) using EKSMP method with  $(k,\ell)= (1,50)$ and KS solver without Cholesky preconditioning.}
    \label{tab:eksmp_convergence}
\end{table}

\newpage
\bibliographystyle{model1-num-names}
\bibliography{Biblio}

\end{document}